\documentclass[11pt,a4paper,english]{article}
\usepackage{graphicx}
\usepackage{amssymb,latexsym}
\usepackage{amsmath}
\usepackage{epsf}
\usepackage{pdfsync}
\usepackage{epsfig}
\usepackage[parfill]{parskip}
\usepackage[matrix,arrow,color]{xy}
\usepackage[usenames]{color}
\usepackage{hyperref}

\usepackage{mathrsfs}

\input{xy}
\xyoption{all}

\input{epsf}

\makeatletter

\catcode`@=11
\renewcommand{\section}
{\@startsection{section}{1}{0pt}{\medskipamount}{\medskipamount}{\large\bf}}
\makeatletter\renewcommand{\subsection}
{\@startsection{subsection}{2}{\z@}{-3.25ex plus -1ex minus -.2ex}
{1.5ex plus .2ex}{\it }}

\numberwithin{equation}{section}
\catcode`@=12

\newcommand{\ba}{\begin{eqnarray*}}
\newcommand{\ea}{\end{eqnarray*}}
\newcommand{\ban}{\begin{eqnarray}}
\newcommand{\ean}{\end{eqnarray}}

\newcommand{\Tr}{{\rm Tr\,}}

\newcommand{\IC}{\mathbb{C}}

\newcommand{\cI}{{\cal I}}

\newcommand{\cN}{{\cal N}}
\newcommand{\cM}{{\cal M}}

\newcommand{\cH}{{\cal H}}

\newcommand{\cE}{{\cal E}}
\newcommand{\cO}{{\cal O}}

\newcommand{\cQ}{{\cal Q}}

\newcommand{\cZ}{{\cal Z}}
\newcommand{\cL}{{\cal L}}
\newcommand{\cF}{{\cal F}}
\newcommand{\cG}{{\cal G}}
\newcommand{\cD}{{\cal D}}

\newcommand{\cP}{{\cal P}}

\newcommand{\sfA}{{\mathsf{A}}}

\newcommand{\sfQ}{{\mathsf{Q}}}

\newcommand{\sfR}{{\mathsf{R}}}

\newcommand{\scrA}{{\mathscr{A}}}
\newcommand{\scrP}{{\mathscr{P}}}

\newcommand{\scrM}{{\mathscr{M}}}

\newcommand{\scrN}{{\mathscr{N}}}

\newcommand{\NDT}{{\tt NC}}
\newcommand{\DT}{{\tt DT}}

\newcommand{\simeqrot}{\reflectbox{\rotatebox[origin=c]{90}{$\simeq$}}}

\newcommand{\mbf}[1]{{\boldsymbol {#1} }}
\newcommand{\complex}{{\mathbb C}} 
\newcommand{\zed}{{\mathbb Z}} 
\newcommand{\real}{{\mathbb R}} 
\newcommand{\torus}{{\mathbb T}}

\def\e{{\,\rm e}\,}

\newcommand{\ch}{{\rm ch}}

\def\ii{{\,{\rm i}\,}}
\def\dd{{\rm d}}

\newcommand{\Hom}{\mathrm{Hom}}

\newcommand{\Ext}{\mathrm{Ext}}

\newcommand{\rank}{\mathrm{rank}}

\newcommand{\Gammaw}{{\widehat\Gamma}}

\def\beq{\begin{equation}}
\def\bee{\begin{equation}}
\def\eeq{\end{equation}}
\def\bea{\begin{eqnarray}}
\def\eea{\end{eqnarray}}
\def\bd{\begin{displaymath}}
\def\ed{\end{displaymath}}

\newcommand{\Cint}{\int\kern-10.5pt-\kern7pt}

\newcommand{\PP}{{\mathbb{P}}}

\newcommand{\be}{\begin{equation}}
\newcommand{\ee}{\end{equation}}

\newcommand\fverbit{\egroup\item[\fbox{\unhbox\pippobox}]}
\newbox\pippobox

\def\a{\alpha}

\def\b{\beta}

{

\def\pa{\partial}

\def\g{\gamma}

\def\bC{{\mathbb C}}

\def\w{\wedge}

\def\be{\begin{equation}}
\def\ee{\end{equation}}
\def\bea{\begin{eqnarray}}
\def\eea{\end{eqnarray}}

\begin{document}

\begin{titlepage}
\setcounter{page}{1}

\begin{center}

\vspace*{1cm}

{\Huge  Defects in Cohomological Gauge Theory and \\[8pt] Donaldson-Thomas Invariants}

\vspace{15mm}

{\large\bf Michele Cirafici}
\\[6mm]
\noindent{\em Center for Mathematical Analysis, Geometry, and Dynamical Systems \\ Departamento de Matem\'atica and LARSyS \\ Instituto Superior T\'ecnico\\
Av. Rovisco Pais, 1049-001 Lisboa, Portugal}\\ Email: \ {\tt cirafici@math.ist.utl.pt}

\vspace{15mm}

\begin{abstract}
\noindent

Donaldson-Thomas theory on a Calabi-Yau can be described in terms of a certain six-dimensional cohomological gauge theory. We introduce a certain class of defects in this gauge theory which generalize surface defects in four dimensions. These defects are associated with divisors and are defined by prescribing certain boundary conditions for the gauge fields. We discuss generalized instanton moduli spaces when the theory is defined with a defect and propose a generalization of Donaldson-Thomas invariants. These invariants arise by studying torsion free coherent sheaves on Calabi-Yau varieties with a certain parabolic structure along a divisor, determined by the defect. We discuss the case of the affine space as a concrete example. In this case the moduli space of parabolic sheaves admits an alternative description in terms of the representation theory of a certain quiver. The latter can be used to compute the invariants explicitly via equivariant localization. We also briefly discuss extensions of our work to other higher dimensional field theories.

\end{abstract}

\end{center}
\end{titlepage}

\newpage

\tableofcontents

\section{Introduction}

The relationship between quantum field theory and geometry has been long and fruitful. In the recent years the BPS sector of supersymmetric field theories has played an increasingly important role and has revealed itself full of surprises. Often a good idea to study this sector is to consider a simplified version of quantum field theory which only contains the relevant information; for example a topological, or better cohomological, gauge theory. Cohomological gauge theories are obtained from supersymmetric gauge theories via the topological twist procedure. The net effect of the twist is to localize the gauge theory onto the moduli space of solutions of the BPS equations, dropping all the perturbative fluctuations. The observables of the cohomological theory can be expressed in terms of the intersection theory on this moduli space. Often this space has an intrinsic geometrical characterization and is associated with an interesting mathematical problem. The cohomological gauge theory becomes a powerful tool to study this problem, via the computational and conceptual insights offered by quantum field theory. One famous example is Donaldson-Witten theory, which arises from the topological twist of $\cN=2$ supersymmetric Yang-Mills, and its relation with the Donaldson invariants. The latter characterize differential four manifolds and correspond to certain integrals over the instanton moduli space of the gauge theory \cite{Witten:1988ze}. On the other hand the low energy dynamics of $\cN=2$ Yang-Mills is governed by the Seiberg-Witten solution \cite{Seiberg:1994rs}, whose equations provide a simpler and equivalent perspective on four manifolds \cite{Witten:1994cg}. 

Another example is the cohomological gauge theory which arises from the twist of maximally supersymmetric Yang-Mills in six dimensions. This theory is expected to capture certain aspects of Donaldson-Thomas theory of Calabi-Yau manifolds. This has been explicitly shown in the case of ordinary Donaldson-Thomas invariants on toric Calabi-Yau varieties \cite{MNOP,Iqbal:2003ds,Cirafici:2008sn,Cirafici:2008ee}. Indeed in the case of toric Calabi-Yau varieties one can use the powerful techniques of equivariant localization to perform explicit computations of the enumerative invariants. These techniques allow to reduce the instanton counting problem of a topological field theory on four or six dimensional toric varieties, to a combinatorial problem \cite{MNOP,Iqbal:2003ds,Cirafici:2008sn,Cirafici:2008ee,neklocal,Gasparim:2008ri,Cirafici:2009ga}. Instanton counting problems have seen dramatic progress in the recent years with the work of Nekrasov \cite{Nekrasov:2002qd,Nekrasov:2003rj}. The basic idea is to localize the instanton measure with respect to the toric action on the instanton moduli space induced by the toric action on the physical space where the gauge theory is defined. The localization formula reduces difficult integrals over the instanton moduli space to a sum over toric fixed points with determined weights. Usually toric fixed points admit a combinatorial classification, in terms of Young diagrams and plane partitions, or generalization thereof. See \cite{Cirafici:2012qc,Szabo:2009vw,Szabo:2011mj} for a recent review within the present context.

The relation between gauge theory and Donaldson-Thomas theory is powerful enough to survive certain generalizations. In particular one can still use instanton counting techniques to study noncommutative Donaldson-Thomas invariants \cite{szendroi}, defined on noncommutative crepant resolutions of Calabi-Yau singularities. In this case the central ingredient is an instanton quiver which governs the local dynamics of the theory on the instanton moduli space. The whole formalism of instanton counting can be adapted to instanton quivers. Instanton quivers seem to be very general structures, with applications to the theory of motivic Donaldson-Thomas invariants and the theory of quantum cluster algebras \cite{Cirafici:2010bd,Cirafici:2011cd}. 

The geometrical structures associated with topological gauge theories are enriched by the presence of defects. Defects can be seen as certain physical modifications of the Feynman integral where for example the relevant fields are assumed to have a prescribed behavior along the defect. In four dimensional Yang--Mills theory the most studied cases are the Wilson and 't Hooft line defects. Surface defect are a higher dimensional generalization of line defects. Depending on the perspective taken, sometimes defects can be thoughts of as path integral insertions and are also customarily called line or surface operators. A certain class of surface defects was throughly studied by Gukov and Witten in the context of the topological twist of $\cN=4$ super Yang--Mills which describes aspects of the geometric Langlands program \cite{Gukov:2006jk,Gukov:2008sn}. These are co-dimension two defects on which the S-duality of $\cN=4$ super Yang-Mills acts non trivially and are mathematically described by parabolic Higgs bundles. These defects also exists in the case of $\cN=2$ super Yang-Mills which is more close to the spirit of this paper, where they have an interesting wall-crossing behavior and provide a deep connection with hyperholomorphic geometry and Hitchin systems \cite{Gaiotto:2009fs,Gaiotto:2011tf}. Surface defects in topological Yang-Mills were further used in \cite{Tan:2009qq,Tan:2010dk} to provide physics proofs of several results concerning the geometry of four manifolds, pointing out tantalizing new connections between invariants of four manifolds and the theory of embedded surfaces \cite{kron1,kron2}. Surface defects in four dimensional gauge theories have been studied from the point of view of instanton counting \cite{Alday:2010vg,Wyllard:2010vi,Wyllard:2010rp,Kozcaz:2010yp,Kanno:2011fw} and topological strings \cite{Kozcaz:2010af,Dimofte:2010tz,Awata:2010bz}.

The purpose of this paper is to lay the foundations of a theory of divisor defects in higher dimensional topological gauge theories and hopefully foster discussion between physicists and mathematicians. Our main playground will be the six dimensional gauge theory associated with Donaldson-Thomas theory on a generic Calabi-Yau, but we will also discuss how our arguments generalize to other topological theories. Geometrically our results suggest the existence of a generalization of Donaldson-Thomas theory and the associated enumerative problem. We will study in detail the modifications needed in the gauge theory to incorporate the defects. These amount roughly to prescribe certain boundary conditions for the gauge field nearby a defect. We then study how to incorporate generalized instanton configurations in the theory coupled to a defect. This can be done in full generality from the gauge theory perspective by defining an appropriate instanton moduli space. The new instanton moduli space is essentially the moduli space of solutions of the Donaldson-Uhlenbeck-Yau equations with prescribed boundary conditions along the divisor. In ordinary Donaldson-Thomas theory one obtains a better behaved moduli space by relaxing the concept of holomorphic bundle to torsion free sheaves. Similarly we argue that in our problem we should consider the moduli space of torsion free sheaves with a parabolic structure along the divisor. This suggest the existence of a generalization of Donaldson-Thomas invariants related to the intersection theory of the moduli space of stable parabolic sheaves on a Calabi-Yau threefold. 

After describing the modifications due to the defect in full generality, we turn to an explicit example and consider the gauge theory on $\complex^3$ with a divisor defect. In this case we can use toric localization techniques to evaluate explicitly the partition function and the new invariants. To do so we construct explicitly the relevant moduli space of parabolic sheaves. It turns out that this moduli space can be identified with a fixed locus of the ordinary moduli space of torsion free sheaves, with respect to a certain discrete action determined by the defect. We introduce an appropriate instanton quiver to study this fixed locus and carry out explicitly the localization computation.

Finally we end with a discussion about possible generalization to other topological field theories, as well as to the more intricate case of defects coupled to lower dimensional defects.

This paper is organized as follows. In Section \ref{surfaceop} we review few results on the theory of surface defects in four dimensional supersymmetric gauge theories, which will be generalized in the rest of the paper. In particular we emphasize the relation between surface defects and instanton counting. In Section \ref{coho} we review some aspects of Donaldson-Thomas theory and its relation with cohomological gauge theory. In Section \ref{divisor} we lay the foundations of a theory of divisor defects in Donaldson-Thomas theory, from the gauge theory point of view. Firstly we discuss the general structure of a divisor defect and afterwards we study the necessary modifications to Donaldson-Thomas theory to include the defect. We do this step by step, by starting with the gauge theory intuition as a guide and finally conjecture that the relevant moduli space to study involves torsion free sheaves with a parabolic structure along a divisor. This suggests the existence of new enumerative invariants associated with the intersection theory of this moduli space. The discussion so far is quite abstract and therefore in Section \ref{instanton} we review the connection between Donaldson-Thomas theory and instanton counting, introducing the necessary ingredients to construct an explicit example in Section \ref{parabolic}. In this example we study Donaldson-Thomas theory with a divisor defect on the affine space and construct explicitly the instanton moduli space. It turns out that this moduli space can be identified with the fixed locus of the moduli space of torsion free sheaves, with respect to a certain discrete action. In Section \ref{defects} we use explicitly this identification to compute the generating function of Donaldson-Thomas invariants with a divisor defect, by generalizing known instanton counting techniques. In Section \ref{defectsH} we briefly discuss divisor defects in other higher dimensional cohomological field theories. We conclude by summarizing our findings and with a discussion about possible generalizations. 

\section{Surface operators in four dimensional $\cN=2$ Yang--Mills} \label{surfaceop}

In this paper we will be interested in co-dimension two defects. In the context of four dimensional gauge theories, these correspond to surfaces, and can be defined via a certain modification of the functional integral which specifies the behavior of all the fields along a two dimensional surface. The class of defects we are interested in has been introduced in the context of $\cN=4$ gauge theories by Gukov and Witten \cite{Gukov:2006jk} as a tool to study certain aspects of the geometric Langland program and has been subsequently generalized to $\cN=2$ gauge theories \cite{Gaiotto:2009fs}. They are sometimes called surface operators, and we will use the term ``defect" and ``operator" interchangeably. We will briefly review the $\cN=2$ case since it is similar to our problem and will motivate some of our conjectures. For simplicity consider topological $\cN=2$ Yang--Mills with a simple gauge group $G$ defined on a certain four manifold $M$. We follow the conventions of \cite{Gukov:2006jk}. The gauge field $A$ is a connection on a principal $G$--bundle $E \longrightarrow M$, and takes values in the real algebra of $G$ (and is therefore considered anti-Hermitian). The covariant derivative is $\dd_A = \dd + A$ and the field strength $F_A = \dd_A^2 = \dd A + A \wedge A$. Consider a (real) co--dimension two surface $\Sigma$ embedded in $M$. Defining a surface operator amounts in prescribing a certain singular behavior for the gauge field restricted to the normal bundle to the surface $\Sigma$. In this case we can assume that locally $M= \Sigma \times K$ where $K$ is the local fiber of the normal bundle, parametrized by $z = r \e^{\ii \theta}$. The gauge field near the surface operator looks like
\begin{equation} \label{surface}
A = \alpha \, \dd \theta + \cdots \ ,
\end{equation}
and in particular, since $\dd \theta = \ii \dd z / z$ the connection is singular as $z \longrightarrow 0$, and the dots stand for non singular terms. Here the parameter $\alpha$ is what specifies the type of surface operator and takes values in the Cartan subalgebra $\frak t = \rm{Lie} \, \torus_G$ of the Lie algebra $\frak g = {\rm Lie} \, G$, where $\torus_G$ is the maximal torus of $G$. To be more precise, the correct gauge invariant concept is the monodromy $\e^{- 2 \pi \alpha}$ of the connection $A$ around a circle of constant radius $r$ \cite{Gukov:2006jk}. Therefore $\alpha$ really takes value in the quotient $\frak t / \Lambda = \torus_G$ where $\Lambda$ is the cocharacter lattice of $G$ \cite{Gukov:2006jk}. To compute the curvature at the origin of the singular connection, one uses that $\dd (\alpha \, \dd \,  \theta) = \alpha  \, \delta_\Sigma$ where $\delta_\Sigma$ is a two form Poincar\'e dual to the surface $\Sigma$. Therefore
\begin{equation}
F_A = 2 \pi  \alpha \, \delta_\Sigma + \cdots \ ,
\end{equation}
and we say that the theory is ``ramified". Note that we are still free to shift $\alpha$ by a lattice vector. This corresponds to the fact that because of the singularity along $\Sigma$, the $G$--bundle $E$ is only defined in a complement of $\Sigma$  in $M$ and we are free to pick an extension of $E$ to all of $M$. Different extensions correspond to different lifts of $\alpha$ from $\frak t / \Lambda$ to $\frak t$. One extension is mapped into another by the gauge transformation $(r , \theta) \longrightarrow \e^{\theta u}$ where $u \in \frak t$ is such that $\e^{2 \pi u} = 1$. Therefore there exists a natural $\torus_G$--bundle over $\Sigma$, since this gauge transformation acts trivially on $\torus_G$. In plain words the gauge field $A$ restricted to $\Sigma$ is a connection on this $\torus_G$-bundle. 

In the most generic case, the surface $\Sigma$ might be non trivially embedded, corresponding to a non vanishing
\begin{equation}
\Sigma \cap \Sigma = \int_M \delta_\Sigma \wedge \delta_\Sigma \ .
\end{equation}
In this case the parameter $\alpha$ has to satisfy appropriate conditions. For example if $G=U(1)$, $\alpha$ is constrained by
\begin{equation}
\int_\Sigma \frac{F_A}{2 \pi} = \alpha \, \Sigma \cap \Sigma \ \in \zed \ ,
\end{equation}
and for general $G$ the analog statement holds in its maximal torus $\torus_G$.

A convenient way to look at supersymmetric field configurations in the presence of a surface operator is to pick an extension and consider $\alpha$ as an element of the Cartan subalgebra $\frak t$. Therefore the connection $A$ is now defined over all of $M$ and its field strength $F$ takes values in $\frak t$ when restricted to $\Sigma$. One considers the bundle $E'$ defined over all of $M$ whose field strength is
\begin{equation}
F_A' = F_A -  2 \pi \alpha \, \delta_\Sigma \ .
\end{equation}
The field strength $F_A'$ is the natural object to consider in the action and in the equations of motion \cite{Gukov:2006jk}; this is equivalent to look for solutions $F_A$ of the BPS equations with the prescribed singularity along the surface operator $\Sigma$. Note that when other fields are involved, as is the case for example for $\cN=4$ Yang--Mills, they should all be subject to an analogous prescription \cite{Gukov:2006jk}. The type of surface operator we have been discussing so far is known as a \textit{full} surface operator, where the parameter $\alpha$ are generic in the torus $\torus_G$. A more general situation is possible and indeed surface operators were classified in  \cite{Gukov:2006jk} in terms of pairs $(\alpha, L)$ where $L$ is a subgroup of $G$ of Levi type, or Levi subgroup for short. This is a subgroup of $G$ whose elements commute with $\alpha$, and which obviously contains $\torus_G$. Indeed $\torus_G$ is a minimal Levi subgroup. A surface operator of type $L$ is defined as a surface operator where $\alpha$ is invariant under the action of $L$ \textit{and} in the functional integral we divide by gauge transformations which take values in $L$ when restricted to $\Sigma$. For example, if $G=SU(r)$ a surface operators is full if $L = U(1)^{r-1}$ and called \textit{simple} if $L = SU(r-1) \times U(1)$.

An equivalent and sometime more useful description of surface operators is in terms of parabolic groups. There is a correspondence between subgroups of $G$ of Levi type and parabolic subgroups of $G_{\complex}$. Given a surface operator whose singularity is parametrized by $\alpha$ we can define the associated parabolic subalgebra $\frak p$ of $\frak g_{\complex}$ spanned by elements $x$ which obey
\begin{equation}
[ \alpha , x] = \ii \lambda \, x \ , \qquad \ \text{with} \ \lambda \ge 0 \ .
\end{equation}
The corresponding group $P \subset G_{\complex}$ is called a parabolic subgroup. Roughly speaking we can think alternatively of the surface operator as a flat connection whose monodromy along $\Sigma$ is determined by the data $(\alpha , L)$ or as a stable holomorphic $G_{\complex}$-bundle whose structure group is reduced to a parabolic subgroup $P$ along $\Sigma$. The equivalence between these two points of view was proven in \cite{mehta}. We will return to this point in the following sections and discuss its generalization to the higher dimensional case. 

As we have already mentioned, topological Yang-Mills provides valuable information about the geometry of four manifolds. Donaldson invariants are defined via integrals over the instanton moduli space. It is natural to consider how this picture gets modified when a surface defect is introduced. In the case of $SU(2)$ the problem has been set up by Kronheimer and Mrowka in \cite{kron1,kron2} and discussed from the field theory point of view in \cite{Tan:2009qq,Tan:2010dk}. The resulting theory is very rich and provides new tools to study the geometry of four manifolds, such as new ``ramified" Donaldson invariants. These are defined in terms of the intersection theory of the moduli space of anti-self dual connections modulo gauge transformations on $E'$ restricted to $M \setminus \Sigma$, the moduli space of ramified instantons. In the language of topological Yang-Mills, ramified Donaldson invariants arise as topological observables, defined via integration over the moduli space $\cM'$ of gauge inequivalent configurations satisfying
\begin{equation} \label{surfinst}
(F'_A)^+ = \left( F_A -  2 \pi \alpha \, \delta_\Sigma \right)^+ = 0 \ .
\end{equation}
Ramified anti-self dual connections are labelled topologically by the instanton number $k$ and a set of monopole numbers $\frak{m}^I$ associated with the $U(1)$ factors of the Levi subgroup $L$. These enter the problem via a coupling to a set of two dimensional theta angles $\eta$ via the term $\Tr \eta \, \frak{m}$ where
\begin{equation}
\Tr \eta \, \frak{m} = \frac{1}{2 \pi} \int_{\Sigma} \Tr \eta \, F_A \ .
\end{equation}
Therefore one can construct the moduli space of ramified instantons $\cM'_{k, \frak{m}}$ and study its intersection theory. In this paper we will initiate an analogous program for Donaldson-Thomas theory of Calabi-Yau threefolds.

The study of the intersection theory of the instanton moduli space in four dimensional gauge theory is notoriously a difficult problem. In the case of toric geometries several technical problems can be overcome by using localization techniques. The toric setting is a natural preliminary step in the study of instanton moduli spaces in the presence of a surface operators. In this case one is interested in the equivariant intersection theory with respect to the toric action on the moduli space, constructed from the toric action on the bulk and the natural toric action associated with the Cartan subalgebra of the gauge group. The simplest four dimensional toric geometry is the affine space $\complex^2$. The natural toric action of $\complex^2$ was used in \cite{Nekrasov:2002qd,Nekrasov:2003rj},  generalizing previous works \cite{Moore:1998et,Moore:1997dj}, to localize the instanton measure onto its fixed points and reduce the instanton counting problem to the purely combinatorial problem of counting Young diagrams. For more details we refer the reader to the recent review \cite{Cirafici:2012qc}, whose notation we'll borrow. When the theory is coupled to a surface operator the natural object to study is the generating function of the equivariant integrals
\begin{equation}
\cZ (\epsilon_1 , \epsilon_2 , \mbf{a} ; L, q, \eta) = \sum_{\frak{m}} \, \sum_{k \in \zed} \ \e^{\Tr \eta \, \frak{m}} \ q^k \ \oint_{\cM'_{k,\frak{m}}} \ 1 \ .
\end{equation}
A conjecture to compute this partition function was proposed in \cite{Alday:2010vg} using the results of \cite{feigin}. This proposal was extensively checked and studied in \cite{Wyllard:2010vi}--\cite{Tan:2013tq}. Consider $\complex^2 [z_1 , z_2]$ with toric action $z_i \longrightarrow \e^{\ii \epsilon_i} \, z_i$ for $i=1,2$. Given the relation between surface operators and parabolic bundles, in \cite{Alday:2010vg} it was proposed to identify (a compactification of) the moduli space of instantons  in the presence of a surface defect, with the moduli space of torsion free sheaves with a certain parabolic structure. Therefore one is led naturally to study the equivariant intersection theory of the moduli space of parabolic sheaves. For the case of a full surface operator the relevant moduli space was studied in \cite{feigin}. The natural toric action of $\complex^2$ lifts to this moduli space and all the quantities of interest can be computed via equivariant localization. More precisely the central object is the moduli space of torsion free sheaves on a compactification of $\complex^2$ with a certain parabolic structure at the divisor $z_2=0$. This moduli space is naturally embedded in the moduli space of torsion free sheaves, simply forgetting the parabolic structure, and one can employ the standard instanton counting techniques: fixed points of the toric action are still isolated, classified by Young diagrams and the form of the character of the instanton deformation complex at a fixed point is known explicitly \cite{feigin}. Indeed a much simpler way to compute the partition function was used in \cite{Kanno:2011fw} following the construction of \cite{finkel}. There exists an explicit map between torsion free sheaves with a parabolic structure and $\Gamma$-invariant torsion free sheaves, with $\Gamma$ an appropriate cyclic group. As a result the problem of instanton counting with a surface operator is reduced to studying the $\Gamma$ fixed locus in the instanton moduli space \textit{without} surface operators. This is precisely the approach we will adapt to our case in Sections \ref{parabolic} and \ref{defects}.

\section{Cohomological Yang-Mills and Donaldson-Thomas invariants } \label{coho}

The main focus of this paper will be the six dimensional topological Yang-Mills theory introduced in \cite{Baulieu:1997jx}-\cite{Baulieu:1997nj}. This theory was afterward discussed in the context of Donaldson-Thomas theory and topological string theory in \cite{Iqbal:2003ds,Cirafici:2008sn,Dijkgraaf:2004te}. We will keep the discussion general, albeit later on we will specialize to the case where the theory is in its Coulomb branch and defined on a toric manifold. In this case the powerful techniques of equivariant localization will be used to compute the gauge theory partition function explicitly. We will follow the review \cite{Cirafici:2012qc}.

\subsection{Generalities}

We start by collecting some definitions and setting up some notation. We will mostly work on a Calabi-Yau threefold $X$, that is a complex manifold with K\"ahler form $J$ and with trivial canonical bundle $K_X = \cO_X$. We will denote by $t = B + \ii J$ the complexified K\"ahler form, where $B$ is the background supergravity two form B-field. Some of our considerations will only depend on the K\"ahler structure and not on the Calabi--Yau condition. We will furthermore assume that $X$ has an hermitian metric $g$, which however will not enter explicitly in our computations. Because of the complex structure the de Rham differential decomposes as $\dd = \partial + \overline{\partial}$. Given a complex hermitian vector bundle $(\cE,h)$ with hermitian metric $h$, a connection $A$ will be associated with a covariant differential $d_A$, which splits as $d_A = \partial_A + \overline{\partial}_A$. In local complex coordinates $z^{\mu}$, $\mu=1,2,3$
\begin{equation}
 \overline{\partial}_A = \dd \overline{z}^{\mu} \frac{\partial}{\partial \overline{z}^{\mu}} + \dd \overline{z}^{\overline{\mu}} A_{\overline{\mu}} \ .
\end{equation}
The corresponding curvature will be denoted by $F_A$ and can be decomposed as $F_A = F_A^{(2,0)} + F_A^{(1,1)} + F_A^{(0,2)}$ where $F_A^{(0,2)} = \overline{\partial}_A^2$. An holomorphic vector bundle is characterized by the equation $F_A^{(0,2)} = 0$. The moduli space of holomorphic bundles on a certain variety plays a prominent role in supersymmetric theories. One obtains a better behaved moduli space by requiring the holomorphic bundles to be also $\mu$-stable. We say that a holomorphic bundle $\cE$ is $\mu$--stable if for any sub-bundle $\cE' \subset \cE$ with $\rank \, \cE' < \rank \, \cE$ we have $\mu (\cE') < \mu (\cE)$ where
\begin{equation}
\mu (\cE) = \frac{\deg \, \cE}{\rank \, \cE} \ .
\end{equation}
Similarly one can introduce the notion of semi-stability, where the inequalities are not strict. Here the degree of the bundle $\cE$ is defined as
\begin{equation}
\deg \, \cE = \int c_1 (\cE) \wedge t \wedge t \ .
\end{equation}
Stable holomorphic bundles can be equivalently characterized by the Donaldson-Uhlenbeck-Yau (DUY) equations
\begin{eqnarray} \label{DUY}
F_A^{2,0} &=& 0
  \ , \nonumber\\[4pt]
F_A^{1,1} \w t \w t &=&
l~t \w t \w t \ .
\end{eqnarray}
Here $l$ is proportional to the degree of the gauge bundle. These equations arise naturally in supersymmeric problems as BPS conditions. The generic strategy consists in considering only the first equation to characterize BPS states as holomorphic bundles and mod out by complexified gauge transformations. The full BPS problem is recovered by imposing the $\mu$--stability condition which is equivalent to the second DUY equation.

\subsection{Cohomological Yang-Mills theory in six dimensions}

The problem of studying Donaldson-Thomas invariants on $X$ is essentially a higher dimensional instanton problem. The associated topological gauge theory is a topological version of six dimensional Yang-Mills. The most economical way of thinking about this theory is via dimensional reduction of super Yang-Mills in ten dimension. After the reduction the six dimensional fields are a connection $A$ on the $G$-bundle $\cE \longrightarrow X$, and the $\mathrm{ad} \, \cE$ valued complex one form Higgs field $\Phi$ and the forms $\rho^{(3,0)}$ and $\rho^{(0,3)}$. The fermionic sector is twisted, that is the fermions can be though of as differential forms thanks to the identification between the spin bundle and the bundle of differential forms
\begin{equation}
S (X) \simeq \Omega^{0,\bullet} (X) \ ,
\end{equation}
which holds on any K\"ahler manifold. Overall the fermionic sector comprises sixteen degrees of freedom which are organized into a complex scalar $\eta$, one forms $\psi^{1,0}$ and $\psi^{0,1}$, two forms $\chi^{2,0}$ and $\chi^{0,2}$ and three forms $\psi^{3,0}$ and $\psi^{0,3}$. The bosonic part of the action is
\begin{eqnarray}
S  &=& \frac{1}{2}\, \int_X\, \Tr \left( \dd_A \Phi \wedge * \dd_A
\overline{\Phi} + \big[\Phi \,,\, \overline{\Phi}~\big]^2 +
\big|F_A^{(0,2)} + \overline{\pa}\,_A^{\dagger} \rho\big|^2 +
\big|F_A^{(1,1)}\big|^2  \right) \nonumber\\  &&+\,
\frac{1}{2} \frac{1}{(2 \pi)^2}\, \int_X \,\Tr \Big(   F_A \w F_A \w t +
\mbox{$\frac{\lambda}{3 \cdot 2 \pi}$}\, F_A
\w F_A \w F_A \Big) \ ,
\label{bosactionTGT}\end{eqnarray}
where $\dd_A=\dd+A $ is the gauge-covariant derivative, $*$ is the Hodge operator with respect to the K\"ahler metric of $X$, $F_A=\dd A+A\wedge A$ is the gauge field strength. Furthermore $\lambda$ is a coupling constant which in a stringy treatment of Donaldson-Thomas theory should be thought of as the topological string coupling constant.
The gauge theory has a BRST symmetry and hence localizes onto the moduli space $\scrM^{\rm inst}_r(X)$  of solutions of the ``generalized instanton"
equations
\begin{eqnarray} \label{inste} F_A^{(0,2)} &=&
  \overline{\pa}\,_A^{\dagger} \rho
  \ , \nonumber\\[4pt]
F_A^{(1,1)} \w t \w t + \big[\rho\,,\, \overline{\rho}\,\big] &=&
l~t \w t \w t \ , \nonumber\\[4pt] \dd_A \Phi &=& 0 \ .
\end{eqnarray}
On a Calabi-Yau we can restrict our attention to minima such that $\rho = 0$. In this case the first two equations on (\ref{inste}) reduce precisely to the Donaldson-Uhlenbeck-Yau equations (\ref{DUY}) and BPS states correspond to stable holomorphic vector bundles. In the following, unless explicitly stated otherwise, we will only consider bundles $\cE$ such that $l=0$. In the string theory picture this corresponds to the counting of D0-D2-D6 brane bound states without D4 brane charge. Furthermore to obtain a better behaved moduli space, we will allow for more general configurations corresponding to torsion free coherent sheaves, as is customary in instanton counting problems (and reviewed for example in \cite{Cirafici:2012qc}); we will however sometimes switch to the more familiar holomorphic bundle language to aid intuition. The moduli space of torsion free coherent sheaves $\scrM^{\rm inst}_r(X)$ stratifies into connected components with fixed characteristic classes. We will denote these components by $\scrM^{\rm inst}_{n , \beta ; r} (X)$ where $\left( \ch_3 (\cE) , \ch_2 (\cE) \right) = (n , -\beta)$.

The local geometry of the moduli space is captured by the instanton deformation complex
\begin{equation} \label{defcomplex}
\xymatrix@1{0 \ar[r] & \Omega^{0,0} ( X , \mathrm{ad}\, \cE)
\ar[r]^{\hspace{-1.3cm} C} & ~\Omega^{0,1} ( X , \mathrm{ad}\, \cE
) \oplus \Omega^{0,3} ( X , \mathrm{ad}\, \cE)
\ar[r]^{\hspace{1.3cm} D_A} & \Omega^{0,2} ( X , \mathrm{ad}\, \cE) \ar[r] & 0 } \ ,
\end{equation}
where $\Omega^{\bullet,\bullet} (X , \mathrm{ad}\, \cE)$ denotes
the bicomplex of complex differential forms taking values in the
adjoint gauge bundle over $X$, and the maps $C$ and $D_A$
represent a linearized complexified gauge transformation and the
linearization of the first equation in (\ref{inste}) respectively. The complex is elliptic; its first cohomology is the Zariski tangent space to the moduli space $\scrM^{\rm inst}_{n , \beta ; r} (X)$ at a certain point, represented by an holomorphic bundle $\cE$ with given characteristic classes. The second cohomology is the normal or obstruction bundle $\scrN_{n,\beta;r}$ which is associated with the kernel of the conjugate operator $D_A^{\dagger}$. The cohomology in degree zero is associated with reducible connections, and will be henceforth assumed to be vanishing. The gauge theory is topological and its partition function reduces to a sum over topological sectors of integrals over the instanton moduli space, with an appropriate measure. This measure is given by the Euler class of the normal bundle ${\rm eul} (\scrN_{n,\beta;r})$. At least formally one has
\begin{equation} \label{ZDT}
\cZ^{DT}_X (q, Q ; r) = \sum_{k \ \beta} \ q^k \ Q^{\beta}  \ \int_{\scrM^{\rm inst}_{n , \beta ; r} (X)} \ {\rm eul} (\scrN_{n,\beta;r}) \ .
\end{equation}
The notation is as follows: We consider $\beta$ as an element of $H_2 (X , \zed)$ and expand it in a basis as $\beta = \sum_i n_i S_i$ with $n_i \in \zed$ and $i=1, \dots , b_2(X)$. Then $Q^{\beta} := \prod_i \, Q_i^{n_i} $ with $Q_i = \e^{- t_i}$ and $t_i = \int_{S_i} \, t$. 

These integrals can be defined more precisely by using a more sophisticated formalism, and correspond to the Donaldson-Thomas invariants. In this paper we will refrain from trying to give them a more mathematically precise meaning in full generality and continue to use the gauge theory intuition. Note that in principle the rank $r$ can be taken arbitrary, corresponding to a $U(r)$ gauge theory on the worldvolume of a stack of coincident D6 branes. However at present we only know how to make computational progress when $X$ is a toric manifold and the gauge theory is in the Coulomb branch, where the gauge symmetry is broken down to the maximal torus $U(1)^r$. In this case the integrals representing Donaldson-Thomas invariants can be defined rigorously and computed explicitly via equivariant localization. We will return to this case later on. We stress however that, at least formally, the gauge theory perspective can be used to study higher rank invariants.

\subsection{Donaldson-Thomas invariants}

The rank $r=1$ case plays a special role in the theory of the topological string. Physically it corresponds to counting bound states of a single D6 brane with a gas of D0-D2 branes. Mathematically it corresponds to the enumerative problem of counting subschemes $Y \subset X$ with fixed topological data. From this point of view the relevant moduli space is the Hilbert scheme of points and curves $\mathrm{Hilb}_{n,\beta} (X )$ with fixed
\begin{equation}
\chi (\cO_Y) = n \ , \qquad [Y] = \beta \in H_2 (X , \zed) \ .
\end{equation}
Equivalently we can consider the moduli space of ideal sheaves. A ideal sheaf $\cI_{Y}$ is a torsion free sheaf with trivial determinant and is associated with a scheme $Y$ via the short exact sequence
\begin{equation} \label{idealseq}
\xymatrix@1{
0 \ar[r] &  \cI \ar[r]^a & \cO_X \ar[r]^b & \cO_Y \ar[r] & 0 \ ,
}
\end{equation}
which simply means that the ideal sheaf $\cI_{Y}$ is the kernel of the restriction map $\cO_X \longrightarrow \cO_Y$. We will denote by $\cI_{n,\beta} (X )$ the moduli space of ideal sheaves on $X$. We define the abelian Donaldson-Thomas invariants as
\begin{equation} \label{DTdef}
\DT_{n,\beta} (X) = \int_{[\cI_{n,\beta} (X )]^{\mathrm{vir}}} 1 \ ,
\end{equation}
using the virtual fundamental class defined in \cite{thomas}. 

In the rank one case there is a certain case which could be regarded as an ``avatar" of our construction, which will be exposed in the next Sections. Given a divisor $D$ in $X$ one can define \textit{relative} Donaldson-Thomas invariants \cite{MNOP2} via integration over the moduli space $\cI_{n,\beta} (X \setminus D )$ of stable ideal sheaves on $X$ relative to $D$. Then relative Donaldson-Thomas invariants are defined, as above, via integration over the virtual fundamental class constructed out of this moduli space. It is not clear to us if this problem and the study of Donaldson-Thomas rank one invariants in the presence of a divisor defect are equivalent.

In this paper we are most interested in the nonabelian problem, where divisor operators are naturally defined. The above definitions don't extend immediately to the nonabelian problem, the main reason being the issue of stability. Roughly speaking when setting up the Donaldson-Thomas problem in this language we are only caring about the holomorphic condition in the DUY equations (\ref{inste}) modulo complexified gauge transformations. As a result in general the moduli space is bigger than it should and the correct moduli space is recovered by selecting only $\mu$-stable sheaves. In the higher rank case it is not known how to do this systematically, while rank one ideal sheaves are automatically $\mu$-stable. This problem persists when the theory is defined with a defect. The only explicit computations that we will be able to carry out explicitly in the presence of a divisor defect, will be in the Coulomb branch of the gauge theory, where the relevant configurations are direct sums of abelian solutions and therefore the stability problem will be sidestepped. In the more general case a certain stability condition should be imposed, as we will discuss in the next Section.

\section{Divisor defects and Donaldson-Thomas theory} \label{divisor}

In this section we will define and study divisor defects in Donaldson-Thomas theory. We will begin by following the gauge theory perspective and define a divisor defect by specifying a certain behavior for the gauge field along a divisor. This corresponds to a modification of the quantum path integral, since now it has to be performed only over those field configurations obeying the prescribed behavior. We then discuss the instanton moduli space when the theory has a defect. Mathematically this means that we only consider those holomorphic connections obeying the required boundary conditions. We later propose to identify this moduli space as the moduli space of parabolic bundles (or better parabolic torsion free sheaves) on $X$ with a certain parabolic structure along the divisor which specifies the divisor operator. The cohomological gauge theory naturally suggest the existence of an enumerative problem associated with this moduli space. We will start to discuss heuristically what kind of enumerative problem we expect from the point of view of the cohomological gauge theory and using the language of holomorphic bundles. Finally we will reformulate the problem more precisely in terms of parabolic sheaves. Since not all the results of this section are rigorously established, we will conclude by summarizing our conjectures.

\subsection{Gauge theory and divisor defects}

We will try now to define operators analogous to the surface operators in $\cN=2 $ Yang-Mills theory in four dimensions. The two main characteristics of surface operators are the fact that they determine a monodromy in the gauge connection by prescribing a singular behavior for the fields along the defect, and the fact that they are classified by Levi subgroups of the gauge group. In extending these concepts to Donaldson-Thomas theory we will try to keep these two characteristics (keeping also in mind that as one moves up in the number of dimensions it is natural to expect more room for extra parameters). Some of our arguments are straightforward extensions of the four dimensional case. 

Consider a $G$-bundle $\cE \longrightarrow X$. To generalize a surface defect to the higher dimensional case we need to consider solutions of the field equations with a prescribed monodromy. To impose a monodromy on the gauge connection we need a co-dimension two defect. The natural object which replaces a surface defect is what we will call a divisor defect. Locally our space has the form $D \times C$ where $D$ is a divisor and $C$ is the local fiber of the normal bundle. We parametrize the gauge connection restricted to $C$ as
\begin{equation} \label{Adiv}
A = \alpha \, \dd \theta + \cdots \ ,
\end{equation}
where we write the local coordinate on $C$ as $z = r \e^{\ii \theta}$ and the dots refer to less singular terms. In this way the singularity is at the origin of $C$. Globally we will require that at each point of the normal plane to the divisor $D$ the gauge connection looks like (\ref{Adiv}). The situation is precisely as in the four dimensional case and the same line of arguments can be extended to conclude that the parameter $\alpha$ takes values in the lattice $\frak t / \Lambda \equiv \torus_G$. The cocharacter lattice is defined as $\Lambda = \pi_1 (\torus_G) = \Hom (U(1) , \torus_G)$. Again $\dd (\alpha \, \dd \, \theta) = \alpha \, \delta_D$, where $\delta_D$ is now a two form Poincar\'e dual to the divisor $D$, and
\begin{equation}
F _A= 2 \pi \alpha \, \delta_D + \cdots \ .
\end{equation}
While these formulas are formally similar to the analogous formulas for a surface operator, one should always keep in mind that $D$ is not a surface but a divisor; this is consistent when going back to four dimensions since in that case a surface has obviously co-dimension two. We will often think of $\alpha$ as an element of $\frak t$. Indeed, as we have already discussed, $\alpha$ can be lifted from $\torus_G$ to $\frak t$ albeit in a non unique fashion. Because of the singularity, the bundle $\cE$ is only naturally defined on $X \setminus D$, but can be extended on the whole of $X$. Each lift of $\alpha$ is associated to a different extension. While there is no natural extension of $\cE$ as a $G$-bundle, $\cE$ can be naturally extended to a $\torus_G$-bundle over $D$. This means that the curvature $F_A$ defined over $X$ is $\frak t$-valued when restricted to $D$. Similarly in the functional integral, we divide by those gauge transformations which are $\torus_G$-valued when restricted to $D$. The two points of view, working on $X \setminus D$ or over all of $X$ with an extension, are complementary and rooted in the field theory description of line defects; for example when talking about (electric) Wilson lines is natural to integrate the gauge field over the defect,  while (magnetic) 't Hooft line defects are more naturally defined by excising the line from the bulk space and specifying boundary conditions for the fields.
 
This is strictly true for a full divisor operator, but as in the four dimensional case we can have more general defects where the maximal torus $\torus_G$ is replaced by a more general Levi subgroup $L \subset G$ that contains $\torus_{G}$. In this case we pick parameters $\alpha \in \frak t$ which commute with $L$ (and we say that $\alpha$ is $L$-regular). We define a divisor operator of type $L$ by prescribing a singular behaviour of the type (\ref{Adiv}) along a certain divisor $D \subset X$ with the requirement that the parameters $\alpha \in \frak t$ are invariant under $L$ \textit{and} in the path integral we divide by the group of gauge transformations which are $L$-valued when restricted to $D$. Now the structure group of $\cE$ restricted to $D$ is $L$ and likewise the connection $F_A$ extended to all of $X$ is $L$-invariant when restricted to $D$.

As we have already explained the pair $(L , \alpha)$ determines a parabolic subgroup of $G$. The simpler example is when $L=\torus_{G}$ in which case the associated parabolic group is called a Borel subgroup $B$ and is the group of upper triangular matrices of appropriate rank. An alternative way of thinking about parabolic subgroups of a group $G$ is as stabilizers of flags on $\complex^n$ (or a generic field). A flag is a sequence of subspaces
\begin{equation}
0 \subset U_1 \subset U_2 \subset \cdots \subset U_n = \complex^r \ .
\end{equation}
In this case $G$ acts on the flag as
\begin{equation}
g \ \left( U_1, \, \dots , U_n \right) = \left( g \, U_1 , \, \dots , g \, U_n \right) \ .
\end{equation}
A flag is said to be \textit{complete} if $n=r$ and $\dim U_i = i$. Complete flags are the flags which are stabilized by Borel subgroups. Indeed the standard complete flag is obtained by choosing
\begin{equation}
U_i = \complex \, e_1 \oplus \complex \, e_2 \oplus \cdots \oplus \complex \, e_i \ ,
\end{equation}
that is the span of the first $i$ elements of the natural basis of $\complex^r$. This perspective will be helpful in the next sections.

We can associate to a divisor defect a parabolic bundle. This correspondence generalizes the four dimensional result for surface defects. This result is based on the correspondence constructed in \cite{mehta} which associates to every flat $G$-bundle with prescribed monodromy $A = \alpha \, \dd \theta + \cdots$ around a point $p$ of a curve $C$, a stable holomorphic $G_{\complex}$-bundle whose structure group is reduced to a parabolic subgroup $P$ at the point $p$. This correspondence is one to one. The results of \cite{mehta} were generalized to higher dimensional varieties in \cite{bhosle}, whose argument we will briefly streamline. Since a divisor has co-dimension two, a divisor defect can still be described by a flat bundle with prescribed monodromy around the defect. A flat connection can be characterized by giving its holonomies along a basis of non-contractible cycles. This is equivalent to giving a representation $\zeta$ of the fundamental group $\pi_1 (X \setminus D) $. The results of \cite{bhosle} show how to associate to a unitary irreducible representation $\zeta$ a stable parabolic bundle. The idea is to take a ``point-wise"  approach and afterwards extend it to a global structure by gluing patches together. If we pick a point $x \in D$, a neighborhood of $x$ in the normal bundle $N$ of $D$ in $X$, restricted to $N \setminus D$, will look like a product of disks, one of which is punctured at the position of the defect. Call this neighborhood $N'_x$. Because of the puncture, $\pi_1 (N'_x) = \zed$ and let $\gamma_0$ be its generator. To construct a parabolic bundle, one firstly considers a representation $\zeta'$ which is the restriction (obtained via the canonical inclusion) of $\zeta$ to $\pi_1 (N \setminus D)$. In particular set $\zeta' (\gamma_0) = \e^{\tau}$ and consider the associated vector bundle $E_{\zeta'}$. The eigenvalues of $\tau$ are called \textit{parabolic weights}. Indeed for simplicity we will assume that $\tau$ is diagonal and of the form
\begin{equation}
\tau = \left( \begin{matrix}  
a_1 \ {\mbf 1}_{r_1} & 0 & & \cdots & 0 \\
0 & a_2 \ {\mbf 1}_{r_2} & 0 & \cdots & 0 \\
\vdots & & \ddots & & \vdots \\
0 & & \cdots & & a_m  \ {\mbf 1}_{r_m} 
\end{matrix}  \right) \ .
\end{equation}
The parameters $0 \le a_1 < \cdots < a_m < 1$ correspond precisely to the parameters $\alpha$ up to a normalization and taking into account their multiplicities. This matrix gives a parabolic structure by determining a certain flag, the one stabilized by it. The construction of \cite{bhosle} shows how to trivialize $E_{\zeta'}$ and how to extend it over to puncture, to a neighborhood $N_x$ of the normal bundle $N$ of $D$ in $X$. Furthermore, as $x$ varies in $D$, the same $\tau$ can be used to define all the extensions to $N_x$, in such a way that the extensions agree on the overlaps of two neighborhoods. The resulting bundle $E(\zeta)$ is defined over all of $X$ and has a parabolic structure on $D$. Since by assumption $\zeta$ is irreducible unitary, $E(\zeta)$ is stable.

In the following we will conjecture a much stronger result, that the generalized instanton moduli space in the presence of a defect (that is, solutions of the Donaldson-Uhlenbeck-Yau equations on the complement $X \setminus D$) can be identified with the moduli space of parabolic bundles (or sheaves) on $X$. To our knowledge even the analog result for ordinary instantons in the presence of surface defects on a generic four manifold and for a generic group $G$, has not been established rigorously in full generality.

\subsection{An example}

Having established the connection between divisor operators and parabolic bundles in abstract terms, we can unpack a bit these formulas and discuss a concrete example. Consider a Levi subgroup of $SU(4)$ of the form
\begin{equation}
\left( 
\begin{matrix}
\bullet & \bullet & 0 & 0 \\
\bullet & \bullet & 0 & 0 \\
0 & 0 & \bullet & 0 \\
0 & 0 & 0 & \bullet 
\end{matrix}
\right) \ ,
\end{equation}
and a parameter $\alpha = \ii \mathrm{diag} ( 2 \varkappa ,2  \varkappa , - \varkappa , -3 \varkappa  )$ with $\varkappa$ a real positive number which we can assume suitably normalized (recall that in our conventions, in a unitary representation of $G$ the gauge field is represented by an anti-hermitian matrix). In the following we adapt an argument of \cite{Gukov:2006jk} to our case. The $\overline{\partial}_A$ operator can be explicitly written near $z=0$ as
\begin{equation}
\overline{\partial}_A = \dd \overline{z} \left( \frac{\partial}{\partial \overline{z}} 
\left( \begin{matrix} 1 & 0 & 0 & 0 \\
0 & 1 & 0 & 0 \\
0 & 0 & 1 & 0  \\
0 & 0 & 0 & 1 
\end{matrix} \right) 
- \frac{\varkappa}{2 \overline{z}} \left( \begin{matrix} 2 & 0 & 0 & 0 \\
0 & 2 & 0 & 0 \\
0 & 0 & -1 & 0  \\
0 & 0 & 0 & -3 
\end{matrix} \right) 
\right) \ .
\end{equation}
This operator can also be written as
\begin{equation}
\overline{\partial}_A = f \, \overline{\partial} \, f^{-1} \ ,
\end{equation}
where $\overline{\partial} = \dd \overline{z} \, \frac{\partial}{\partial \overline{z}}$ and
\begin{equation}
f = \left( \begin{matrix}
(\overline{z} \, z)^{\varkappa} & 0 & 0 & 0 \\
0 & (\overline{z} \, z)^{\varkappa} & 0 & 0 \\
0 & 0 & (\overline{z} \, z)^{-\varkappa/2} & 0 \\
0 &  0 & 0 & (\overline{z} \, z)^{-3 \varkappa/2}
\end{matrix}
\right) \ .
\end{equation}
Now, consider an holomorphic section of the adjoint bundle $\mathrm{ad} \, E$ associated with the principal $SU(4)$ bundle $E$. Given an $\mathrm{ad} \, E$ valued function $s$, the condition that it is also an holomorphic section is that $\overline{\partial}_A \, s = 0$, which implies also $\overline{\partial} \left( f^{-1} \, s \, f \right) \equiv \overline{\partial} \, \tilde{s} = 0$. This gives a set of partial differential equations which are solved by
\begin{equation} \label{umatrix}
s = \left(
\begin{matrix}
u_{11} & u_{12} & u_{13} \, (z \overline{z})^{\frac32 \varkappa} & u_{14} \, (z \overline{z})^{\frac52 \varkappa} \\
u_{21} & u_{22} & u_{23} \, (z \overline{z})^{\frac32 \varkappa} & u_{24} \, (z \overline{z})^{\frac52 \varkappa} \\
u_{31} \, (z \overline{z})^{-\frac32 \varkappa} & u_{32} \, (z \overline{z})^{-\frac32 \varkappa} & u_{33} & u_{34}  \, (z \overline{z})^{\varkappa} \\
u_{41}  \, (z \overline{z})^{- \frac52 \varkappa} & u_{42}  \, (z \overline{z})^{- \frac32 \varkappa} & u_{43} \, (z \overline{z})^{-\varkappa} & u_{44}  
\end{matrix}
\right) \ .
\end{equation}
Here $u_{ij}$ are arbitrary holomorphic functions which we assume obey a suitable condition to make $s$ traceless. Regularity at $z=0$ implies that $u_{3i} (0)$ for $i=1,2$ and $u_{4i} (0)$ for $i=1,2,3$ must vanish. In particular this means that at the position of the divisor operator $\tilde{s}$ has the form (recall that $\tilde{s} = f^{-1} \, s \, f$)
\begin{equation}
\left(
\begin{matrix}
\bullet & \bullet & \bullet & \bullet \\
\bullet & \bullet & \bullet & \bullet \\
0 & 0 & \bullet & \bullet \\
0 & 0 & 0 & \bullet 
\end{matrix}
\right) \ .
\end{equation}
and we see explicitly the reduction of the structure group to a parabolic subgroup at the position of the divisor operator. This is, as we have explained, the holomorphic analog of the monodromy associated with the flat connection (\ref{Adiv}). We stress that the resulting parabolic subgroup depends rather sensitively on the actual values of the elements of $\alpha$, as is clear from the form of (\ref{umatrix}).

\subsection{Generalized instanton moduli spaces}

So far we have discussed divisor defects in general. We will now describe the generalized instanton moduli space when the gauge theory is defined in the presence of a defect. This will set up the stage for the study of Donaldson-Thomas invariants in the presence of a defect. However we will for the time being continue with the gauge theory language and return to the more abstract problem in the following. From the gauge theory perspective we would like to study gauge connections on $X \setminus D$ which have the form (\ref{Adiv}) near $D$, and where the non-singular terms correspond to a generalized instanton. To begin with consider a rank $r$ hermitian vector bundle $\cE$ defined over all of $X$ and with Chern character $\ch (\cE)$. We will firstly consider the case when the divisor defect is full. If this is the case, the bundle $\cE$ decomposes in a closed tubular neighbor to the divisor $D$ as a sum of line bundles
\begin{equation} \label{decompE}
\cE \vert_D = \bigoplus_{i=1}^r \cL_i \ .
\end{equation}
In other words we have an abelian gauge theory on $D$. This gauge theory will be characterized by the topological charges associated with the line bundles $\cL_i$. We can therefore use the Chern classes $c_1 (\cL_i)$ to parametrize the reduction of $\cE$ on $D$. As we will see later on, the abelian gauge theory on $D$ has certain ``theta-angles" associated with the integrals of the Chern classes $c_1 (\cL_i)$. For future convenience we define
\begin{eqnarray} 
m_i &=& \int_D \ \ch_2 (\cL_i) \ , \nonumber \\
h_i &=& \int_{D \cap D} \ c_1 (\cL_i) \ ,  \label{bundlered}
\end{eqnarray}
and form the vectors $\frak m = (m_1 , \cdots , m_r)$ and $\frak h = (h_1 , \cdots , h_r)$ which characterize the reduction of $\cE$ over $D$. Note that for a line bundle,  the Chern character $\ch_2 (\cL_i)$ is completely specified by $c_1 (\cL_i)$; this will not be true anymore when we will consider torsion free sheaves. The need to specify the parameters $\frak h$ might seem strange at first, but will become very natural when we will study the action of the gauge theory.

Consider now the more general case of a divisor defect characterized by a pair $(L , \alpha)$. In this case the structure group of the bundle $\cE$ when restricted to $D$ is $L$. The subgroup $L$ of $G$ will have generically  abelian factors and non-abelian factors. For example, if $G$ has rank $r$, a next-to-maximal subgroup of Levi type is always isomorphic to $SU(2) \times U(1)^{r-1}$ or a $\zed_2$ quotient thereof \cite{Gukov:2006jk}. Since the structure group is reduced on $D$ the bundle $\cE$ will split in a neighbor of $D$ as
\begin{equation}  \label{decompEL}
\cE \vert_D = \bigoplus_{i=1}^e \cE_i^{(L)} \ ,
\end{equation}
where the notation stresses that the decomposition depends on the Levi subgroup $L$ and $e$ is the number of factors. For example in the case of a next-to-maximal Levi subgroup all the factors except one are line bundles. Now to parametrize the reduction of the structure group along $D$ we can use the Chern characters $\ch (\cE_i^{(L)})$ and, for later convenience, we introduce the topological numbers
\begin{eqnarray} 
m_i &=& \int_D \ \ch_2 (\cE_i^{(L)}) \ , \nonumber \\
h_i &=& \int_{D \cap D} \ c_1 (\cE_i^{(L)}) \ . \label{bundleredL}
\end{eqnarray}
We will  use same notation as above, $\frak m = (m_1 , \cdots , m_e)$ and $\frak h = (h_1 , \cdots , h_e)$. The full case is recovered when $e=r$. Note that the first Chern classes associated with the traceless subgroups vanish.

To define the generalized instanton moduli space in the presence of a defect, we can take two points of view, by working on the complement of $D$ or by working over all of $X$. We will now describe these two perspectives. Firstly we define the affine space of connections on the complement of $D$, following the arguments of \cite{kron1,kron2}. We pick a reference connection $A^0$ on $\cE$, such that $A^0 \vert_D$ respects the  decomposition (\ref{decompEL}) over $D$. Now we define a connection $\tilde{A}^{(\alpha)}$ on $\tilde{\cE} = \cE_{X \setminus D}$ as
\begin{equation}
\tilde{A}^{(\alpha)} = A^0 + \alpha \ \dd \theta \ .
\end{equation}
The notation $\tilde{(\bullet)}$ will be reserved for quantities $(\bullet)$ defined over $X \setminus D$. The affine space of connections modeled on $\Omega^{1} \left(X \setminus D ;  \mathrm{ad} \,  \tilde{\cE} \right)$ is defined as
\begin{equation}
\tilde{\scrA}^{(\alpha)} \left( L ; X \setminus D \right)  = \left\{ \, \tilde{A} = \tilde{A}^{(\alpha)} + \tilde{a} \ \vert \  \ \tilde{a} \in \Omega^{1} \left(  X \setminus D ;  \mathrm{ad} \,  \tilde{\cE} \right) \right\} \ .
\end{equation}
In other words $\tilde{a}$ is a smooth connection valued in the restriction of the bundle $\cE$ to $X \setminus D$. Similarly we can define the moduli space 
\begin{equation} \label{Mbundle1}
\tilde{\scrM}^{(\alpha)} \left( L ; X \setminus D \right) = \left. \left\{ \tilde{A} \in \tilde{\scrA}^{(\alpha)} \left( L ; X \setminus D \right)  \, \Big{\vert} \,  \begin{matrix} \tilde{F}_{\tilde{A}}^{(0,2)} = 0 , \\[4pt]  \tilde{F}_{\tilde{A}}^{(1,1)} \w t \w t = 0 \end{matrix} \right\} \right/ \mathcal{\tilde{G}}
\end{equation}
of critical points of the DUY equations in (\ref{inste}) in the complement of the divisor $D$ (recall that we require that the field $\rho$ vanishes identically). Here $\mathcal{\tilde{G}} = \mathrm{Aut} (\tilde{\cE})$ is the group of gauge transformations on $X \setminus D$.

As emphasized in \cite{Gukov:2006jk,Tan:2009qq} a more convenient point of view is to lift $\alpha$ from $L$ to $\frak t$ and deal with bundles defined over all $X$ with a prescribed reduction along $D$. In other words we will now consider $G$-bundles over $X$ whose structure group is reduced to $L$ along $D$. More practically this amounts in dealing with field strengths $F'_A = F_A - 2 \pi \alpha \, \delta_D$ and the moduli problem defined by the equations
\begin{eqnarray} \label{instdiv} 
(F_A   -  2 \pi \alpha \ \delta_D)^{(0,2)} &=&
  \overline{\pa}\,_A^{\dagger} \rho
  \ , \nonumber\\[4pt]
(F_A   -  2 \pi \alpha \ \delta_D)^{(1,1)} \w t \w t + \big[\rho\,,\, \overline{\rho}\,\big] &=&
l~t \w t \w t \ 
\ .
\end{eqnarray}
where again we are only interested in solutions with $\rho=0$. In this case these equations become a direct generalization of (\ref{surfinst}). Note that in principle one can study them also when $\rho$ is non vanishing, specifying an appropriate behavior for $\rho$ near the defect. In this way one could define a more general class of divisor defects. Although interesting, we will not pursue this problem in this paper. We will further assume $l=0$. The correct way to think about the equations (\ref{instdiv}) is that the source $\delta_D$ is forcing the gauge field strength to obey the desired boundary condition on $D$. From this perspective the proper moduli space is
\begin{equation} \label{Mbundle2}
\scrM^{(\alpha)} \left( L ; X \right) = \left. \left\{ A \in  \scrA (X)  \, \Big{\vert} \ \begin{matrix}  
(F_A   -  2 \pi \alpha \, \delta_D)^{(0,2)} = 0 , \\[4pt]
(F_A   -  2 \pi \alpha \, \delta_D)^{(1,1)} \w t \w t = 0 \end{matrix} \
\right\} \right/ \mathcal{G}_D \ ,
\end{equation}
where $\scrA (X)$ is the affine space of connections modeled on $ \Omega^{0,1} \left( X ;  \mathrm{ad} \,  \cE  \right)$ and $\mathcal{G}_D$ is now the group of gauge transformations which take value in $L$ along $D$. When we want to stress the topological numbers of $\cE$ we will use the notation $\scrM^{(\alpha)}_{n, \beta, u ; r} \left( L ; X \vert \{ \ch (\cE^{(L)}_i) \} \right)$, where $(n , -\beta, u)  = (\ch_3 (\cE) , \ch_2 (\cE), c_1(\cE))$,  or $\scrM^{(\alpha)}_{n, \beta, u ; r}$ for brevity. 

Note that nothing guarantees that for generic choices of $X$, $D$ and $L$, these moduli spaces are non-empty.  Furthermore, we will see later that there are other moduli spaces which are more natural. We will argue that, precisely as it is done in ordinary instanton problems, a nicer moduli space can be obtained by using torsion free sheaves. The moduli spaces (\ref{Mbundle1}) and (\ref{Mbundle2}) will then be replaced by moduli spaces of parabolic sheaves. For the time being, we will still use the gauge theory language and discuss the gauge theory action.

\subsection{Action and quantum parameters}

The cohomological gauge theory associated with the moduli problem (\ref{instdiv}) has an effective action which involves the new field strength $F_A' = F_A - 2 \pi \alpha \, \delta_D$. This action depends explicitly on $\alpha$ and so it would seem that the theory depends explicitly on the extension of the bundle over $D$. However this is not the case, since this dependence can be eliminated via an appropriate shift of the ``theta-angles". We will now show this explicitly. In particular in the action we will have to deal with integrals over all of $X$ containing the delta function form $\delta_D$, Poincar\'e dual to $D$. Integrals of this form can be dealt with easily; if we denote by $i \, : \, D \longrightarrow X$ the embedding of $D$ in $X$, then for a generic four form $\varpi$
\begin{equation}
\int_X \ \varpi \wedge \delta_D = \int_D i^* \, \varpi \ .
\end{equation}
To compute these integrals we will use Chern-Weil theory since it is easier to keep track of the factors of $\alpha$, and in particular we will leave the pullback $i^*$ implicit to simplify the notation.

The relevant terms from the action are the topological densities, since the rest of the action is a BRST variation. These can be read from (\ref{bosactionTGT}) and are
\begin{eqnarray} \label{FFF}
\int_X \Tr F_A' \wedge F_A' \wedge F_A' &=& \int_X \Tr F_A \wedge F_A \wedge F_A - 6 \pi \int_D \Tr \alpha \, F_A \wedge F_A + 3 (2 \pi)^2 \int_{D \cap D} \Tr \alpha^2 \, F_A \nonumber  \\[4pt]  && - (2 \pi)^3 \Tr \alpha^3 \, D \cap D \cap D \ ,
\\[4pt] \label{FFk}
\int_X \Tr F_A' \wedge F_A' \wedge t &=& \int_X \Tr F_A \wedge F_A \wedge t - 4 \pi \int_D \Tr \alpha \, F_A \wedge t + (2 \pi)^2 \Tr \alpha^2 \, \int_{D \cap D} t  \ ,
\\[4pt] \label{Fkk}
\int_X \Tr F_A' \wedge t \wedge t &=& \int_X \Tr F_A \wedge t \wedge t - 2 \pi \,  \Tr \alpha \, \int_D t \wedge t \ .
\end{eqnarray}
Note that the last term in (\ref{FFF}), the last term in (\ref{FFk}) and the last term in (\ref{Fkk}) are field independent and can be dropped from the functional integral with no harm. We see that the six dimensional topological action is naturally coupled to a set of theta-angles. In particular the instanton action is sensitive to integrals over $D \cap D$, explaining why we choose the parametrization (\ref{bundleredL}). Using this parametrization we can write 
\begin{eqnarray}
\frac{1}{(2 \pi)^2} \, \frac12  \int_D \Tr \alpha \, F_A \wedge F_A  &=&   \alpha_i \ m^i \ , \nonumber \\[4pt]
\frac{1}{2 \pi} \, \int_{D \cap D} \Tr \alpha^2 \, F_A &=&  \alpha_i^2 \ h^i \ , \nonumber \\[4pt]
\frac{1}{2\pi} \, \int_D \Tr \alpha \, F_A \wedge t^D &=& \alpha_i \, o^i_a \, t^D_a \ , \label{thetaintegers}
\end{eqnarray} 
where we have used the fact that in each Levi subfactor, $\alpha$ can be seen as a diagonal matrix proportional to the identity, and that the K\"ahler form $t$ restricts to the K\"ahler form $t^D$ of $D$. The parameters $o^i_a$ arise by expanding $F_A$ and $t^D$ in a basis of 2-cycles. All the extra terms (\ref{thetaintegers}) will contribute to the functional integral with terms of the form 
\begin{equation}
q^{- \alpha_i \, m^i + \mbox{$\frac12$} \alpha_i^2 \, h^i} \ \prod_{a=1}^{b_2 (D)} \ Q_{D , a}^{- o_a^i \, \alpha_i} \ ,
\end{equation}
with $Q_{D , a} = \e^{- t_a^D}$.

The four dimensional gauge theory on the divisor defect will have its own quantum parameters. To simplify the discussion let us assume that the divisor operator is full, that is the structure group $G$ is reduced precisely to its Cartan $\torus_G$. If $G$ has rank $r$ then $\torus_G$ is isomorphic to $U(1)^r$. For each rank 1 factor, the gauge theory on $D$ is abelian and given in terms of a line bundle $\cL$ on $D$. In this case we have the freedom to include the term
\begin{equation}
\exp 2 \pi \ii \left( \eta \int_{D} \ch_2 (\cL) + \gamma \int_{D} c_1 (\cL) \wedge k + \sigma \int_{D \cap D} c_1 (\cL) \right) 
\end{equation}
in the functional integral. In general we will have additional parameters for each rank $1$ factor; we will regroup them in vectors such as $\eta^I = (\eta^1 , \cdots , \eta^r)$. In the more general case of a divisor operator preserving a Levi group $L$, we will have parameters $\eta^i$, $\gamma^i$ and $\sigma^i$ associated with each subfactor of $L$. Note that the quantum numbers associated with a subfactor can be zero. Therefore the most general term which we can include in the functional integral will be
\begin{equation}
\exp 2 \pi \ii \left( \eta_i \ m^i  + t^{D}_a \ \gamma_i \ o_a^i + \sigma_i \ h^i \right) \ ,
\end{equation}
where  $t_a^{D}$ are the K\"ahler moduli of $D$. We can use this term to absorb any shifts in the parameters $\alpha_i$ which appear in the equations (\ref{FFF}--\ref{Fkk}) via a shift of the four dimensional parameters $\eta$, $\gamma$ and $\sigma$. Therefore the full six dimensional instanton action in the presence of a divisor defect, is independent of $\alpha$, precisely as in \cite{Gukov:2006jk}. Indeed this seems to be a general feature of co-dimension two defects. Recall that the parameters $\alpha$ were lifted from $\torus_G$ to the Cartan subalgebra $\frak t$, each different lift parametrizing different extensions of the bundle $\cE$ on $X \setminus D$ over $D$. Now we see that the ambiguity present in this lift has disappeared from the instanton action altogether. In other words the ambiguity is fixed by quantum effects.

Finally we are left with the integrals of the characteristic classes of the sheaf $\cE$ over $X$. In the following, unless stated otherwise, we will also drop the remaining term in equation (\ref{Fkk}) as is customary in Donaldson-Thomas theory (this term can always be reinstated by shifting the K\"ahler form). 

\subsection{Parabolic sheaves}

In ordinary (generalized) instanton problems the second equation in (\ref{instdiv}) can be traded for a suitable stability condition, at the price of complexifying the gauge group. Similarly flat bundles with the prescribed singularity $A = \alpha \ \dd \theta + \cdots$ are in one to one correspondence with stable parabolic bundles \cite{bhosle}. It is natural to conjecture that this set of ideas holds more generally and that the generalized instanton problem in the presence of a divisor defect corresponds to studying stable holomorphic parabolic $G_{\complex}$-bundles with non trivial characteristic classes, that is bundles such that $\overline{\partial}_A^2 = F^{(2,0)}_A = 0$ and an appropriate reduction of the structure group along $D$. As we have already remarked we can trade the description in terms of the Levi subgroup $L$ of $G$ and the parameter $\alpha$, with the corresponding parabolic subgroup $P$ of $G_{\complex}$. In the ordinary case without the divisor operator, one would work with $\mu$-stable holomorphic bundles. When the defect is present, one needs to define an appropriate stability condition which preserves the parabolic structure. Furthermore we already know that the cohomological gauge theory problem requires us to enlarge the generalized instanton moduli space to include torsion free sheaves. This leads us naturally to the idea that the appropriate moduli space to study is the moduli space of stable parabolic torsion free sheaves. Moduli spaces of this sort have been studied before in the literature \cite{murayama}, although for our purposes we prefer to label the moduli spaces by the characteristic classes of the sheaves separately, and not by their Hilbert polynomial.

While in the context of holomorphic bundles it is natural to talk about the reduction of the structure group to a parabolic subgroup, when dealing with sheaves it is more appropriate to think directly of the parabolic structure in terms of the flag stabilized by the parabolic subgroup $P$. Therefore following \cite{bhosle,murayama} we define a parabolic structure as follows. Consider $\cE$ a torsion free sheaf on $X$. For bundles a parabolic structure means that the structure group is reduced along the divisor. Indeed since we are working over all of $X$, over $D$ we have a natural $L$-bundle specified by the weights $\alpha$; equivalently the associated parabolic subgroup $P$. However, instead of giving the group $P$ at each point of $D$, we could simply use the alternative characterization of the parabolic subgroup $P$ via the flag which is stabilized by $P$. Therefore at each point of $D$ we have a flag of vector spaces. This information glues nicely over all of $D$ since we are, after all, just talking about the structure group of a bundle. In the more general language of sheaves we do not have anymore a structure group to talk of, but we can still specify a flag of subsheaves over $D$. Therefore we \textit{define} a parabolic structure on the sheaf $\cE$ over $D$ as a flag $\cG_{\bullet}$ of subsheaves of $\cE \vert_D  = \cE \otimes \cO_D$
\begin{equation}
 \cE \vert_D = \cG_1 (\cE) \supset \cG_2 (\cE) \supset \cdots \supset \cG_{l} (\cE) \supset \cG_{l+1} (\cE) = 0 \ .
\end{equation}
To this flag we associate also a $l$--tuple of weigths $0 \le a_1 < a_2 < \cdots < a_l \le 1$. In the case of a parabolic bundle, these weights are the eigenvalues of the matrix $\tau$ introduced previously. Up to a (conventional) normalization they coincide with the parameters $\alpha$ (where if two $\alpha_i$ are equal they are associated with a single $a_i$). 

We define the parabolic $\mu$--weight of the sheaf $\cE$ as
\begin{equation}
\mu^{wt} (\cE) = \sum_{i=1}^{l-1} a_i \int \left( c_1(\cG_i (\cE)) - c_1 (\cG_{i+1} (\cE)) \right) \wedge t \wedge t \ ,
\end{equation}
and the parabolic degree of $\cE$ as
\begin{equation}
\frak p \deg (\cE) = \deg (\cE) + \mu^{wt} (\cE) \ .
\end{equation}
Finally we can say that a parabolic sheaf $\cE$ is $\mu$--stable (respectively semistable) if for any subsheaf $\cE'$ of rank $\rank \, \cE' < \rank \, \cE$ one has $\frak p \mu (\cE') < \frak p \mu (\cE)$ (respectively $\frak p \mu (\cE') \le \frak p \mu (\cE)$) where
\begin{equation}
\frak p \mu (\cE) = \frac{\frak p \deg \cE}{\rank \, \cE} \ .
\end{equation}

We will denote by $\scrP_{n, \beta , u ; r}^{(\alpha)} (X , D \vert \{ \ch (\cG_i (\cE)) \} )$ the moduli space of rank $r$,  $\mu$--stable parabolic sheaves $\cE$ on $X$ with a parabolic structure along the divisor $D$ and fixed characteristic classes  $( \ch_3 (\cE) , \ch_2 (\cE)  , c_1 (\cE) = n , -\beta , u)$. We propose that the theory of Donaldson-Thomas invariants in the presence of a divisor operator is the study of the intersection theory of this moduli space. 

There is an alternative definition of parabolic sheaves which will be more convenient in the following, where instead of giving a filtration for the restriction of the sheaf $\cE$ on $D$, one constructs directly a filtration of sheaves over $X$. More precisely, one can define a torsion free parabolic sheaf $\cE$  as a torsion free sheaf with the following parabolic structure over $D$: a filtration
\begin{equation}
\cF_\bullet \ : \ \ \cE = \cF_1 (\cE) \supset \cF_2 (\cE) \supset \cdots \supset \cF_l (\cE) \supset \cF_{l+1} = \cE (-D) \ ,
\end{equation}
together with a sequence of weights $0 \le a_1 < a_2 < \cdots < a_l \le 1$. The two definitions are equivalent and are related by the short exact sequence
\begin{equation} \label{parabSES}
\xymatrix@1{
0 \ar[r] &  \cF_i (\cE)  \ar[r] &\cE \ar[r] & \cE \vert_D / \cG^i (\cE) \ar[r] & 0 \ .
}
\end{equation}
In particular from this short exact sequence it follows the relation between the Chern characters of the sheaves $\cG^i (\cE)$ supported on $D$ and the sheaves $\cF_i (\cE)$
\begin{equation}
\ch (\cF_i (\cE)) = \ch (\cE) - \ch (\cE \vert_D) + \ch (\cG^i (\cE)) \ .
\end{equation}
The reduction of the gauge field due to the defect is simply parametrized by the Chern classes of the sheaves $\cF_i$ via the sequence (\ref{parabSES}). It is therefore natural to consider a moduli space with these characteristic classes fixed. In other words we are interested in parametrizing the moduli space of parabolic sheaves with fixed $\left( \ch_3 (\cF_i (\cE))  , \ch_2 (\cF_i (\cE))  , c_1 (\cF_i (\cE)) \right) = \left( n_i , - \beta_i , u_i \right)$. Note that $\cF_1 = \cE$ whose characteristic classes are $( \ch_3 (\cE) , \ch_2 (\cE)  , c_1 (\cE) = n , -\beta , u)$. We will denote this moduli space with $\scrP_{n , \beta , u ; r}^{ (\alpha)} (X , D \vert \{ \ch (\cF_i (\cE)) \} ) $, or $\scrP_{n , \beta , u ; r}^{ (\alpha)}$ for simplicity.

Note that these moduli spaces can be empty. Furthermore as in ordinary Donaldson-Thomas theory, even if they are non-empty, we don't expect them to be well behaved. In this paper we will make no attempt to resolve this issue. We will simply assume that, as in ordinary Donaldson-Thomas theory one can construct a meaningful intersection theory with more sophisticated tools. Indeed in the following sections we will see an explicit example where this is possible, the case of affine $\complex^3$ with a specific divisor operator, where the relevant moduli space is actually a fixed point set of the moduli space of ordinary torsion free sheaves on $\complex^3$.

\subsection{Summary}

Finally we summarize our conjectures. We have argued that the cohomological gauge theory problem in the presence of a divisor operator reduces to the study of the intersection theory of the moduli space $\scrM^{(\alpha)}_{n , \beta, u ; r}$. In particular the gauge theory provides a natural measure, the Euler class of the normal bundle $\mathrm{eul} (\scrN^{(\alpha)}_{n ,\beta, u ; r})$, the second cohomology of the instanton deformation complex (\ref{defcomplex}), when restricted to configurations obeying (\ref{instdiv}).  As in ordinary Donaldson-Thomas theory we can construct a generating function
\begin{equation}
\cZ^{DT}_{(X , D)} (q, Q ; r) = \sum_{n , \, \beta, \, u} \sum_{\frak{m} , \frak{h} , \frak{o}} \ q^k \ Q^{\beta} \ v^u \ \e^{ 2 \pi \ii \left( \eta^i \, m_i  + t^{D}_a \, \gamma^i \, o_i ^a + \sigma^i \, n_i \right)}
  \ \int_{\scrM^{(\alpha)}_{n, \beta, u ; r} \left( L ; X \vert \{ \frak{m} , \frak{h} , \frak{o}  \} \right)} \ \mathrm{eul} (\scrN^{(\alpha)}_{n ,\beta, u ; r}) \ .
\end{equation}
where we have for convenience parametrized the Chern classes $\{ \ch (\cE^{(L)}_i) \}$ in terms of the integers appearing in (\ref{thetaintegers}) and have omitted the dependence on the theta angles in the partition function $\cZ^{DT}_{(X , D)} (q, Q ; r)$. Note that the form of the instanton action require us to work in a topological sector where the topological numbers $ \left( \frak{m} , \frak{h} , \frak{o} \right)$ are fixed. We have kept track of the first Chern class $u$ introducing a counting parameter $v$; however this can be safely ignored since changing $u$ can be always accomplished by tensoring the sheaves with a line bundle, which does not affect the local structure of the moduli space.

We do not expect this moduli space to be well behaved in any sense; we have argued that it should be replaced with the moduli space of stable parabolic torsion free sheaves on $X$ with a certain parabolic structure along the divisor $D$ (or in principle finitely many divisors in $X$). We do not know how these two moduli spaces are related, hopefully the moduli space $\scrP_{\beta , n, u ; r}^{(\alpha)} $ can be seen as a compactification of the space $\scrM^{(\alpha)}_{n , \beta, u ; r}$ in some appropriate sense. Regrettably we do not have any argument to support this conclusion. Therefore we are led to \textit{define} the problem of Donaldson-Thomas theory with a divisor defect as the study of the moduli space $\scrP_{\beta , n, u ; r}^{(\alpha)} $. While this is certainly better, it is not at all clear that this moduli space has a meaningful virtual intersection theory. We assume that it is the case and consider the associated enumerative problem. In this language Donaldson-Thomas invariants in the presence of a divisor defect should be captured by integrals of the form
\begin{equation}
\DT^{(\alpha)}_{n,\beta, u;r} (X , D \vert  \{ \frak{m} , \frak{h} , \frak{o} \} ) = \int_{\scrP_{\beta , n, u ; r}^{(\alpha)} (X , D \vert \{ \frak{m} , \frak{h} , \frak{o} \} )} \ \mathrm{eul} (\scrN^{(\alpha)}_{n ,\beta, u ; r}) \ .
\end{equation}
Note that strictly speaking we have defined the numbers $ \{ \frak{m} , \frak{h} , \frak{o} \} $ only in the case of vector bundles via Chern-Weil theory, since it easier to keep track of the factors of $\alpha$. However the same definition can be extended to the case of torsion free sheaves. To this end one just needs to introduce a reference bundle $\cP^{(\alpha)}$ over $X$ associated with the connection $A = \alpha \, \dd \theta$. Then the Chern character $\ch (\cE \otimes \cP^{(\alpha)})$ gives the natural generalization of the topological numbers $ \{ \frak{m} , \frak{h} , \frak{o} \} $ in the case where $\cE$ is not a bundle.

The discussion so far has been rather abstract. Of course it would be interesting to make any of these ideas more rigorous. For the time being, to clarify certain aspects and to show an explicit example, we will study in the next Sections the case $X = \complex^3$ with a divisor defect. This case, far from being trivial already in the case without divisor operator, is rather instructive since we will be able to construct the moduli space rather explicitly and to compute directly the partition function via the techniques of equivariant localization with respect to the natural toric action on $\complex^3$.

\section{Instanton counting and Donaldson-Thomas invariants} \label{instanton}

So far we have considered the defect problem in full generality on a generic Calabi-Yau manifold $X$. Now we would like to turn to an example where some explicit computations can be made. This generically is a rather difficult task. As is well known the situation simplifies if the Calabi-Yau is toric (henceforth non-compact). Cohomological gauge theories on a toric manifold are reduced via localization to a diagrammatic evaluation, in term of vertices and propagators,  the latter associated with toric invariant curves \cite{MNOP,Iqbal:2003ds}. The building block of this construction is the vertex, which corresponds to the gauge theory partition function on the affine space. This partition function can be explicitly computed via instanton counting techniques. Therefore, as a preliminary step towards the evaluation of the full partition function on a generic toric Calabi-Yau, we would like to compute the cohomological gauge theory partition function on $\complex^3$ with a defect. To do so we will now briefly review the general instanton counting formalism for Donaldson-Thomas invariants on $\complex^3$ and in the next Sections discuss the modifications which occur in the presence of a defect.

The problem of Donaldson-Thomas invariants on a toric Calabi-Yau can be rephrased as a generalized instanton counting problem. In the case of $\complex^3$ this problem can be solved quite explicitly in the Coulomb branch of the theory. The formalism developed in \cite{Cirafici:2008sn} is based on a generalized ADHM construction which parametrize ideal sheaves on $\complex^3$. This is derived by an explicit homological construction of the moduli space. As a first step one compactifies $\complex^3$ to $\PP^3$ by adding a divisor at infinity and then tries to parametrize the moduli space of torsion free sheaves with fixed characteristic classes on $\PP^3$ and with a trivialization condition on the divisor at infinity. This is done in practice by rewriting each sheaf $\cE$ via a Fourier-Mukai transform whose kernel is the diagonal sheaf of $\PP^3 \times \PP^3$. This procedure is rather technical and the outcome is a certain spectral sequence. Upon imposing some conditions, including the trivialization on a divisor at infinity, the spectral sequence degenerates into a four term complex characterized by a series of matrix equations. These equations give a finite dimensional parametrization of the instanton moduli space and are called generalized ADHM equations. Based on these equations one can construct a certain topological quantum mechanics which can be used to compute the relevant instanton integrals. This quantum mechanics can be though of as arising from the quantization of the collective coordinates around each instanton solution.

In this section we will recall the basis of this construction and show how it can be used to compute instanton integrals. The topological quantum mechanics is given in terms of the homological data of the generalized ADHM construction. The formalism is based on two vector spaces $V$ and $W$ with $\dim_{\complex} V = n$ and $\dim_{\complex} W = r$. Physically $n$ represent the instanton number of the gauge field configuration while $r$ is the rank of the gauge theory. The generalized ADHM formalism can be conveniently described via the auxiliary quiver diagram
\begin{equation} \label{ADHMquiver}
\vspace{4pt}
\xymatrix@C=20mm{
& \ V \ \bullet \ \ar@(ul,dl)_{B_2} \ar@(ur,ul)_{B_1} \ar@(dr,dl)^{B_3}
\ar@{.>}@(ur,dr)^< < < <{\varphi} & \ \bullet \ W \ar@//[l]_{ \ \ \ \ \ I} 
} \ .
\vspace{4pt}
\end{equation}
Recall that a quiver $\sfQ=(\sfQ_0,\sfQ_1)$ is an algebraic entity defined by a set of
nodes $\sfQ_0$ and by a set of arrows $\sfQ_1$ connecting the
nodes. To the arrows one can associate a set of relations
$\sfR$. The \textit{path algebra} of the quiver is defined as the
algebra of all possible paths in the quiver modulo the ideal generated
by the relations; the product in the algebra is the concatenation of
paths whenever this makes sense and $0$ otherwise. This algebra will
be denoted as $\sfA = \complex \sfQ / \langle\sfR \rangle$. A
representation of the quiver $\sfQ$ can be constructed by associating
a complex vector space to each node and a linear map between vector
spaces for each arrow, respecting the relations $\sfR$. Instanton counting is determined in terms of the representation theory of this quiver with certain relations, the generalized ADHM equations. These facts were thoroughly reviewed in \cite{Cirafici:2012qc}.

We have introduced the morphisms
\begin{eqnarray}
(B_1 , B_2 , B_3 , \varphi) \in \Hom_\bC (V,V) \qquad \mbox{and} \qquad
I\in\Hom_\bC(W,V) \ .
\label{bosnaive}\end{eqnarray}
Here $\varphi$ is a finite dimensional analogous of the field $\rho^{(3,0)}$ and we will be mainly interested in representations of the ADHM quiver where $\varphi$ is trivial. The fields $B_{\alpha}$ and $\varphi$ are in the adjoint representation of $U(n)$ while $I$ is a $U(n) \times U(r)$ bifundamental. Furthermore all fields transform under the lift of the natural toric action of $\torus^3$ on $\complex^3$, to the instanton moduli space. Under the full symmetry group $U(n)\times U(r)\times\torus^3$
the transformation rules are
\begin{eqnarray}
B_\a ~&\longmapsto~& \e^{- \ii \epsilon_\a}\, g_{U(n)} \, B_\a \, g_{U(n)}^{\dagger} \ , \nonumber\\[4pt]
\varphi~&\longmapsto~& \e^{- \ii (\epsilon_1 +
\epsilon_2 + \epsilon_3)} \,  g_{U(n)} \, \varphi \,
g_{U(n)}^{\dagger} \ , \nonumber\\[4pt]
I ~&\longmapsto~& g_{U(n)}\, I\, g_{U(r)}^{\dagger}\ .
\end{eqnarray}
The quiver quantum mechanics on $\complex^3$ is characterized by the
bosonic field equations
\begin{eqnarray}
\mathcal{E}_{\a} \,&:&\, [B_\a , B_\b] + \sum_{\g=1}^3\, \epsilon_{\a\b\gamma} \,\big[B_\gamma^{\dagger}
\,,\, \varphi\big] = 0 \ , \nonumber\\[4pt]
\mathcal{E}_\lambda  \,&:&\, \sum_{\a=1}^{3}\, \big[B_\a \,,\,
B_\a^{\dagger}\,\big] + \big[\varphi \,,\, \varphi^{\dagger}\,\big] + I\,
I^{\dagger} = \varsigma \ , \nonumber\\[4pt] \mathcal{E}_I \,&:&\,
I^{\dagger} \,\varphi = 0 \ ,
\label{quiverdefeqs}\end{eqnarray}
where $\varsigma>0$ is a Fayet--Iliopoulos parameter. 

Starting from these equations and the ``matter content" (\ref{bosnaive}), one can use the standard formalism of topological field theories explained in \cite{Moore:1998et,Moore:1997dj} to construct a cohomological matrix quantum mechanics model to study the moduli space of solutions of these equations. Roughly speaking one starts from the BRST transformations
\be
\begin{array}{rlllrl}
\cQ \, B_\a &=& \psi_\a & \quad \mbox{and} & \quad \cQ \, \psi_\a& = \ [\phi , B_\a] -
\epsilon_\a\,
B_\a \ , \\[4pt] \cQ \, \varphi& =& \xi & \quad \mbox{and} & \quad
\cQ \, \xi &= \ [\phi ,\varphi] - (\epsilon_1 + \epsilon_2 + \epsilon_3) \varphi \ , \\[4pt]
\cQ \, I &=& \varrho & \quad \mbox{and} & \quad \cQ \, \varrho& = \
\phi \,I - I \,\mbf a \ ,
\end{array}
\label{BRSTmatrices}\ee
where $\phi$ is the generator of $U(n)$ gauge transformations and $\mbf a = \mathrm{diag} (a_1 , \dots , a_r)$ parametrizes an element of the Cartan subalgebra
$\mathfrak{u}(1)^{\oplus r}$. Finally one needs to introduce Fermi multiplets corresponding to the anti-ghosts and the auxiliary fields associated with the equations (\ref{quiverdefeqs}) as well as the gauge multiplet to close the BRST algebra. This procedure is standard and we refer the reader to the literature; the outcome is that the quiver quantum mechanics path integral localizes onto the fixed points of the BRST charge. These were classified in \cite{Cirafici:2008sn} in terms of $r$-vectors of plane partitions $\vec{\pi} = \left( \pi_1 , \dots , \pi_r \right)$ with $|\vec \pi | = \sum_l |\pi_l|  = k$ boxes. We think of a plane partition as an ordinary Young diagram $\lambda$ together with a ``box piling function" $\pi \, : \, \lambda \longrightarrow \zed_+$ such that $\pi_{i,j} \ge \pi_{i+m,j+n}$ where $m,n \ge 0$. Each partition $\pi_i$ corresponds to a $\torus^3$-fixed ideal sheaf $\mathcal{I}_{\pi_i}$ supported on a $\torus^3$ invariant zero dimensional subscheme in $\complex^3$. Fixed points under the full $\torus^3 \times U(1)^r$ action have the form 
\begin{equation} \label{fixedpointsheaf}
\cE_{\vec \pi} = \cI_{\pi_1} \oplus \cdots \oplus \cI_{\pi_r} \ .
\end{equation}
We take the three torus to be $\torus^3 = (t_1 = \e^{\ii \epsilon_1} , t_2 = \e^{\ii \epsilon_2} , t_3 = \e^{\ii \epsilon_3} )$ and we will denote by $E_l$ the module generated by $e_l = \e^{\ii a_l}$. At a torus fixed point the vector spaces $V$ and $W$ decompose under the $\torus^3 \times U(1)^r$ action as follows
\begin{eqnarray}
V_{\vec\pi} = \sum_{l=1}^r \,e_l~ \sum_{(n_1,n_2,n_3)\in \pi_l}\,
t_1^{n_1-1} \,t_2^{n_2-1}\,t_3^{n_3-1} \qquad \mbox{and} \qquad W_{\vec\pi} =
\sum_{l=1}^r \,e_l \ .
\label{decompos}
\end{eqnarray}

Linearization of $\cE_{\alpha}$ and $\cE_{I}$ around each fixed point leads to the following instanton deformation complex
\begin{equation} \label{adhmdefcomplexC}
\xymatrix{
  \Hom_\complex (V_{\vec\pi} , V_{\vec\pi})
   \quad\ar[r]^{\!\!\!\!\!\!\!\!\!\!\!\!\!\!\!\!\sigma} &\quad
   {\begin{matrix} \Hom_\complex (V_{\vec\pi} , V_{\vec\pi} \otimes Q )
   \\ \oplus \\
   \Hom_\complex (W_{\vec\pi} , V_{\vec\pi}) \\ \oplus  \\ \Hom_\complex (V_{\vec\pi} ,
   V_{\vec\pi}  \otimes \bigwedge^3 
   Q) \end{matrix}}\quad \ar[r]^{\tau} & \quad
   {\begin{matrix} \Hom_\complex (V_{\vec\pi} , V_{\vec\pi}  \otimes \bigwedge^2
       Q) \\ \oplus \\ 
       \Hom_\complex (V_{\vec\pi},W_{\vec\pi} \otimes \bigwedge^3 Q)
   \end{matrix}}
} \ ,
\end{equation}
which is a local model of the instanton moduli space around a fixed point. Here $Q\cong\complex^3$ is a $\torus^3$ module generated by  $t_\a^{-1}=\e^{-\ii\epsilon_\a}$. The equivariant index of this complex compute the virtual sum $\Ext^1_{\cO_{\PP^3}} \ominus \Ext^2_{\cO_{\PP^3}} $ where we are only considering irreducible connections, which corresponds to the assumption that $\Ext^0_{\cO_{\PP^3}}$ vanishes. At a fixed point $\vec\pi$ the equivariant index can be expressed in terms of the characters of the representations at the torus fixed points
\begin{equation} \label{character}
\ch_{\torus^3 \times U(1)^r} \left(  T_{\vec{\pi}}^{\mathrm{vir}}  \scrM^{\rm inst}_{n,0;r}(\IC^3)  \right) = 
W_{\vec\pi}^\vee \otimes V_{\vec\pi} -
{V}_{\vec\pi}^\vee \otimes W_{\vec\pi} + (1-t_1)\, (1-t_2)\,
(1-t_3)~ {V}^\vee_{\vec\pi} \otimes V_{\vec\pi} \ ,
\end{equation}
where we have used the Calabi--Yau condition to set $t_1\,
t_2\, t_3=1$. The enumerative invariants are defined via virtual localization on the instanton moduli space
\begin{equation}
\DT_{n , r} \left( \complex^3 \right) = \int_{[ \scrM^{\rm inst}_{n,0;r}(\IC^3) ]^{\rm vir}} \ 1 = \sum_{[\cE_{\vec{\pi}}] \in \scrM^{\rm inst}_{n,0;r}(\IC^3)^{\torus^3 \times U(1)^r} } \ \frac{1}{\mathrm{eul} \left( T_{\vec{\pi}}^{\mathrm{vir}}  \scrM^{\rm inst}_{n,0;r}(\IC^3)  \right) } \ ,
\end{equation}
where the right hand side takes values in the polynomial ring $\mathbb{Q} [\epsilon_1 , \epsilon_2 , \epsilon_3 , a_1 , \dots , a_r]$ (see for example \cite{Cirafici:2012qc,Szabo:2009vw} for a review within the present context). The virtual tangent space at a fixed point $[\cE_{\vec{\pi}}] \in \scrM^{\rm inst}_{n,0;r}(\IC^3)$ is given by
\begin{equation}
T_{\vec{\pi}}^{\mathrm{vir}}  \scrM^{\rm inst}_{n,0;r}(\IC^3) = T_{\vec{\pi}}\scrM^{\rm inst}_{n,0;r}(\IC^3) \ominus ( \scrN_{n,0;r})_{\vec{\pi}} = \Ext^1_{\cO_{\PP^3}} \left( \cE_{\vec{\pi}} , \cE_{\vec{\pi}} \right)  \ominus \Ext^2_{\cO_{\PP^3}} \left( \cE_{\vec{\pi}} , \cE_{\vec{\pi}} \right) \ .
\end{equation}
As explained in~\cite{Cirafici:2008sn}, the equivariant
index (\ref{character}) computes the ratio of the Euler classes of the obstruction and tangent bundles. It turns out that this ratio is just a sign and in particular independent on the equivariant parameters $\epsilon_i$ and $a_l$. The contribution of an instanton configuration labelled by the $r$-tuple  $\vec{\pi} = \left( \pi_1 , \dots , \pi_r \right)$ is just $(-1)^{r | \vec{\pi} |}$. This allows us to write down the generating function for the Coulomb branch invariants
\begin{equation}
\cZ^{U(1)^r}_{DT} (\complex^3) = \sum_{|\vec{\pi}|} \, q^{|\vec{\pi}|} \ 
\frac{\mathrm{eul}(\scrN_{n,0;r})_{\vec{\pi}}}{\mathrm{eul}\big(T_{\vec{\pi}}\scrM^{\rm inst}_{n,0;r}(\IC^3)\big)} =   \sum_{\vec{\pi}} \ (-1)^{r |\vec{\pi}|} \ q^{|\vec{\pi}|} \ .
\end{equation}
We should stress that this partition function only captures the theory in its Coulomb branch. Furthermore the relevant instanton moduli space was obtained by compactifying $\complex^3$ to $\PP^3$ and imposing a framing condition at infinity. In the following we will use an equivalent compactification of $\complex^3$ to $\PP^1 \times \PP^1 \times \PP^1$ which corresponds to adding a point at infinity to each one of the three $\complex$ spanned by the coordinate $z_i$. Note that on physical grounds it is clear that changing the compactification divisor at infinity, as long as one imposes a trivialization condition, does not changes the moduli space. 

\section{Parabolic sheaves on the affine space and orbifold sheaves} \label{parabolic}

So far we have kept the discussion of divisor operators fairly general. Now, after having reviewed the connection between Donaldson-Thomas invariants and generalized instanton counting, we would like to give a more concrete example by considering a divisor operator on the affine space $\complex^3$. This amounts to consider the moduli space of coherent sheaves on $\complex^3$ with a certain parabolic structure on a divisor. Furthermore, since we will need to use the techniques of virtual localization, we will only consider the gauge theory in its Coulomb branch. This is precisely the situation in four dimensional instanton counting in the presence of a surface operator and we will presently try to generalize it to the case of $\complex^3$. The parabolic structure we choose to impose correspond to a divisor operator located at $z_3=0$ and extended along the two non compact planes parametrized by the coordinates $z_1$ and $z_2$. We will argue that this problem is equivalent to the enumerative problem of $\Gamma$ equivariant ideal sheaves on $\complex^3$, where $\Gamma$ is an appropriate discrete group action which is determined by the type of divisor operator. We expect this to be quite a generic result. The reason is that studying the gauge theory in the complement of the divisor is a similar problem to blowing down the divisor to produce a singularity. This is literally true in the four dimensional case and for certain surface operators \cite{kron1}. When the result of the blow-down is an orbifold singularity, it is natural to expect the orbifold action to select the relevant instanton configurations when the defect is removed. However it is hard to make this connection concrete in general and we will limit ourselves to the affine case. Our construction will be rather explicit.

Our approach is inspired by the analogous construction in four dimensional gauge theories, where (ordinary) instanton counting in the presence of a surface operator is expressed in terms of parabolic sheaves on $\complex^2$ and then reformulated in terms of orbifold sheaves \cite{Kanno:2011fw,feigin}. In the four dimensional case the instanton moduli space is obtained by compactifying $\complex^2$ to $\PP^1 \times \PP^1$ and the presence of a surface operator is induced by imposing a parabolic structure on one of the divisors with $\PP^1$ topology. This is equivalent to a moduli space of torsion free sheaves \textit{without} any parabolic condition but invariant with respect to an appropriate orbifold action; in other words the moduli space of instantons with a surface operator is the $\Gamma$-fixed component of the moduli space of instanton \textit{without} any parabolic structure \cite{finkel}. 

\subsection{Moduli spaces of parabolic sheaves}

In our case we will consider a compactification of $\complex^3$ to $\PP^1 \times \PP^1 \times \PP^1$ and will be interested in coherent sheaves with a certain parabolic structure on a divisor. Since the original space $\complex^3$ is non-compact the proper objects to study are sheaves with a framing condition. When we want to distinguish the three $\PP^1$ we will label them by the projective coordinates as $\PP^1_{z_i}$ where $i=1,2,3$. We will denote by $\cD = \PP^1_{z_1} \times \PP^1_{z_2} \times 0_{z_3}$ the divisor corresponding to the defect, and by $\cD_{\infty} = \PP^1_{z_1} \times \PP^1_{z_2} \times \infty_{z_3} \sqcup \, \PP^1_{z_1} \times \infty_{z_2} \times \PP^1_{z_3} \sqcup \, \infty_{z_1} \times \PP^1_{z_2} \times \PP^1_{z_3}$ the divisor at infinity. We will identify the moduli space of generalized instantons in the presence of a divisor defect as the moduli space of torsion free sheaves with a framing condition on $\cD_{\infty}$, a parabolic structure on $\cD$, and fixed characteristic classes.

Let us be more precise with the definition of our moduli space of parabolic sheaves $\scrP_{\mbf d}$, which is based on \cite{negut,finkel,feigin}. For notational convenience, now we will consider only the case where the divisor operator is associated with the Levi subgroup $\torus_G$ or equivalent a parabolic Borel subgroup $B$, and comment later on how they can be extended to the more general case. Let us fix a $r$-tuple of integers  $\mbf d = (d_0 , \cdots , d_{r-1})$ which will play the role of instanton numbers. 

A parabolic sheaf $\cF_{\bullet}$ is a flag of torsion free sheaves of rank $r$ on $\PP^1 \times \PP^1 \times \PP^1$
\begin{equation}
\cF_{0} (- \cD) \subset \cF_{-r+1} \subset \cdots \subset \cF_{-1} \subset \cF_0  \ .
\end{equation}
We will furthermore require the following conditions
\begin{itemize}
\item[(1)] \textit{Framing}. The sheaves in the flag are locally free on $\cD_{\infty}$, together with a framing isomorphism
\begin{equation} \label{framing}
\xymatrix{
 \cF_{0} (- \cD) \vert_{\cD_{\infty}} \ar[r] \ar[d]^\simeqrot & \cF_{-r+1} \vert_{\cD_{\infty}}  \ar[r] \ar[d]^\simeqrot & \cdots \ar[r] \ar[d]^\simeqrot &   \cF_0 \vert_{\cD_{\infty}}  \ar[d]^\simeqrot \\
\cO^{\oplus r}_{\cD_{\infty}} (-\cD) \ar[r] & W^{(1)} \otimes \cO_{\cD_{\infty}} \oplus \cO^{\oplus r-1}_{\cD_{\infty}} (-\cD) \ar[r] & \cdots \ar[r]  & W^{(r)} \otimes \cO_{\cD_{\infty}}
}
\end{equation}
At infinity $\cF_0$ is isomorphic to the trivial rank $r$ bundle $\cO^{\oplus r}$. By choosing a basis we identify this bundle with the vector space $W^{r} = \langle w_1 , \dots , w_r \rangle$. Similarly $W^{(i)} = \langle w_1 , \dots , w_i \rangle$ for $i = 1 , \dots , r$ are $i$-dimensional vector spaces, and the flag
\begin{equation} \label{flag}
W^{(1)} \subset W^{(2)} \subset \cdots \subset W^{(r)} = W \ ,
\end{equation}
is determined by the parabolic structure (that is, it is the one stabilized by the parabolic subgroup; in the case of a Borel subgroup, which corresponds to a full divisor operator, this is just the standard complete flag).  
\item[(2)] \textit{Chern character}. The framing condition implies $\ch_1 (\cF_k) = k \, [\cD]$ where $[\cD]$ denotes the fundamental class and $-r < k \le 0$ as above. We furthermore require $c_2 (\cF_i) = 0$ and $c_3 (\cF_i) = - d_i$. In other words the Chern classes are specified by the \textit{degree} of the parabolic sheaf  $\mbf d = (d_0 , \cdots , d_{r-1})$ and by the framing condition.
\end{itemize}
We shall denote this moduli space\footnote{Note that technically we are only imposing a quasi-parabolic structure, by specifying a flag. The reason for this is that in the remainder of this paper we are going to use equivariant localization in the Coulomb branch. The weights $a_i$ enter in the definition of $\mu$-stability.  In the Coulomb branch the relevant toric fixed point configurations will be identified with ideal sheaves, and therefore stability will not be an issue. By partial abuse of language we will still talk of ``parabolic" sheaves.} by $\scrP_{\mbf d}$. Later on we will see how this construction gets modified when the divisor operator is not full.

In plain words we want to study torsion free sheaves with a certain parabolic structure and fixed Chern character on a compactification of $\complex^3$. The framing condition assures that the sheaves are trivial at infinity, which corresponds to the need to impose boundary conditions on the physical fields. Note that the sheaves at infinity are twisted and have poles at $z_3= 0$. Therefore for $-r < k \le 0$, the local sections of $\cF_k \vert_{\infty_{z_1} \times \infty_{z_2} \times \PP^1_{z_3}}$ are those local sections of $\cF_0 \vert_{\infty_{z_1} \times \infty_{z_2} \times \PP^1_{z_3}} = W \otimes \cO_{\PP^1_{z_3}}$ which take value in $W^{(k+r)}$ at $0_{z_3}$; this gives a connection with the theory of Laumon spaces, which we will however not pursue in this paper. Different divisor operators will be associated with different parabolic subgroups and therefore to different parabolic structures. As the parabolic structure changes, so does the stabilized flag and therefore the moduli space, as we will see later on. Therefore we have a rule to associate to any divisor operator on $\complex^3$ a different moduli space of torsion free sheaves.  The computation of observables in the gauge theory is reduced to the study of the intersection theory of this moduli space. As in the case without the divisor defect, integration over this moduli space has to be defined carefully. We will sidestep this problem by arguing that this moduli space can be thought of as a fixed component of the generalized instanton moduli space and evaluate the equivariant integrals via virtual localization.

As an aside remark, the moduli space of parabolic sheaves $\scrP_{\mbf d}$ parametrizes sheaves with vanishing $c_2$. Nothing would prevent us to consider a more complicated moduli space with a non vanishing second Chern class. For example we could choose $c_2$ to be the class Poincar\'e dual to one of the $\PP^1$ within $\cD$. Upon imposing appropriate conditions on the moduli space, such as compatibility with the parabolic structure on $\cD$ as well as extra conditions on the sheaves restricted to $\cD$, this would correspond to a surface defect supported on the divisor defect. Or we could consider a second divisor defect, with support at say $\{ z_2 = 0 \}$ which intersect the first one along a surface defect with support on a $\PP^1$. Overall by considering defects within defects or intersecting defects we find a rather rich structure, with additional layers of complexity. In this paper we shall not pursue this interesting direction and hope to resume the discussion elsewhere.

\subsection{Fixed points}

There is a natural action of $\torus^3 \times U(1)^r$ on $\scrP_{\mbf d}$. Recall that the $\torus^3$ fixed points on $\mathrm{Hilb}^n (\complex^3)$ are classified by plane partitions with $n$ boxes, corresponding to ideals $\cI_{\pi}$ with support on a torus fixed point. Similarly $\torus^3 \times U(1)^r$-fixed rank $r$ sheaves have the form $\cE_{\vec \pi} = \cI_{\pi_1} \oplus \cdots \oplus \cI_{\pi_r}$ and are classified in terms of $r$-vectors of plane partitions $\vec \pi = ( \pi_1 , \dots , \pi_r )$. A fixed point in $\scrP_{\mbf d}$ is roughly the same, except for further restrictions imposed by the parabolic structure and the framing condition. The latter implies that certain elements in the fixed points decomposition are twisted by the divisor $\cD$. A fixed point in $\scrP_{\mbf d}$ is a parabolic sheaf $\cF_{\bullet}$ where
\begin{equation}
\cF_{k-r} = \bigoplus_{1 \le l \le k} \ \cI_{\pi_l^{(k)}} \ w_l \ \oplus \bigoplus_{k < l \le r} \ \cI_{\pi_l^{(k)}} \, (- \cD) \ w_l \ .
\end{equation}
Here $\{ \pi_l^{(k)} \}_{1 \le k,l \le r}$ is a collection of plane partitions obeying certain properties that we will outline in a moment. The index $l$ is a Lie algebra index while $k$ refers to the position within the flag. 
Compatibility with the flag structure implies that $\cI_{\pi_l^{(k)}} \subset \cI_{\pi_l^{(k+1)}}$. This implies a corresponding condition on the plane partitions which we will denote by $\pi_l^{(k)} \supset \pi_l^{(k+1)}$. In the previous sections we have denoted a plane partition by the triple $\pi = (m,n, \pi_{m,n})$ where $(m,n)$ correspond to a Young diagram $\mu$. Equivalently we can use the notation $\pi = (\mu, \pi_\mu)$ and denote the Young diagram by $\mu = (\mu_0 \ge \mu_1 \ge \cdots)$, specifying the number of boxes in each row. Then 
\begin{equation}
\pi^{(i)} = (\mu , \pi_\mu) \supset \pi^{(j)} = (\lambda , \pi_{\lambda}) \ \iff \ \mu_i \ge \lambda_i \ \forall \ i \ge 0 \ \text{and} \ \pi^{(i)}_{m,n} \ge \pi^{(j)}_{m,n} \ \forall \ (m,n) \in \mu \ .
\end{equation}
Colloquially, one plane partition lies ``inside" the other (and therefore corresponds to a bigger ideal). Similarly we define
\begin{equation}
\pi^{(i)} = (\mu , \pi_\mu) \supseteq \pi^{(j)} = (\lambda , \pi_{\lambda}) \ \iff \ \mu_i \ge \lambda_{i+1} \ \forall \ i \ge 0 \ \text{and} \ \pi^{(i)}_{m,n} \ge \pi^{(j)}_{m,n+1} \ \forall \ (m,n) \in \mu \ .
\end{equation}
Compatibility with the flag structure therefore requires that the collection $\{ \pi_l^{(k)} \}_{1 \le k,l \le r}$ obeys the following property
\begin{eqnarray} \label{piordering}
& \pi_1^{(1)} \supset \pi_1^{(2)} \supset \cdots \supset \pi_1^{(r)} \supseteq \pi_1^{(1)}  \ , \cr
& \pi_2^{(2)} \supset \pi_2^{(3)} \supset \cdots \supset \pi_2^{(1)} \supseteq \pi_2^{(2)} \ , \cr
& \ \vdots \qquad \vdots \cr
& \pi_r^{(r)} \supset \pi_r^{(1)} \supset \cdots \supset \pi_r^{(r-1)} \supseteq \pi_r^{(r)} \ .
\end{eqnarray}
The presence of the relations $\supseteq$ is due to the twist in the framing condition and to the inclusion $\cF_0 (-\cD) \subset \cF_{-r+1}$ in the flag. Indeed, recall that $\cD$ is described by $\{ z_3 = 0 \}$. A twist by $\cD$ will therefore relate a monomial $z_1^m z_2^{\pi_{m,n}} z_3^n$ with the monomial $z_1^m z_2^{\pi_{m,n}} z_3^{n+1}$. 

To further clarify, consider a full divisor operator in a $U(4)$ gauge theory. A fixed point in $\scrP_{\mbf d}$ is a flag $\cF_0 (-\cD) \subset \cF_{-3} \subset \cF_{-2} \subset \cF_{-1} \subset \cF_0$ where
\begin{eqnarray}
\cF_0 &=& \cI_{\pi_1^{(4)}} \, w_1 \oplus \cI_{\pi_2^{(4)}} \, w_2 \oplus \cI_{\pi_3^{(4)}} \, w_3 \oplus \cI_{\pi_4^{(4)}} \, w_4 \ , \cr 
\cF_{-1} &=& \cI_{\pi_1^{(3)}} \, w_1 \oplus \cI_{\pi_2^{(3)}} \, w_2 \oplus \cI_{\pi_3^{(3)}} \, w_3 \oplus \cI_{\pi_4^{(3)}} (-\cD)  \, w_4 \  ,  \cr 
\cF_{-2} &=& \cI_{\pi_1^{(2)}} \, w_1 \oplus \cI_{\pi_2^{(2)}} \, w_2 \oplus \cI_{\pi_3^{(2)}} (-\cD) \, w_3 \oplus \cI_{\pi_4^{(2)}} (-\cD) \, w_4  \ , \cr 
\cF_{-3} &=& \cI_{\pi_1^{(1)}} \, w_1 \oplus \cI_{\pi_2^{(1)}} (-\cD) \, w_2 \oplus \cI_{\pi_3^{(1)}} (-\cD) \, w_3 \oplus \cI_{\pi_4^{(1)}} (-\cD) \, w_4 \ .
\end{eqnarray}
For example, let us look at the $w_2$ components. From the flag we see immediately that $\pi^{(2)}_2 \supset \pi^{(3)}_2 \supset \pi_2^{(4)}$. Furthermore, since $\cF_0 (-\cD) \subset \cF_{-3}$, we must have that $\cI_{\pi_2^{(4)}}  (-\cD) \subset  \cI_{\pi_2^{(1)}} (-\cD)$, and therefore $\pi^{(4)}_2 \supset \pi_2^{(1)}$. Finally, looking again at the flag, we see that $ \cI_{\pi_2^{(1)}} (-\cD) \subset  \cI_{\pi_2^{(2)}}$ which implies $\pi^{(1)}_2 \supseteq \pi^{(2)}_2$. By collecting all these results we see that it must be that $\pi_2^{(2)} \supset \pi_2^{(3)} \supset \pi_2^{(4)} \supset \pi_2^{(1)} \supseteq \pi_2^{(2)}$, as expected from (\ref{piordering}).

\subsection{Parabolic sheaves as orbifold sheaves}

Up to now we have argued that the correct object to study to understand instanton counting on $\complex^3$ in the presence of a full divisor operator is the moduli space of parabolic sheaves $\scrP_{\mbf d}$. Now we would like to give an alternative description of this moduli space as the fixed component of $\scrM^{\rm inst}_{n , 0 ; r} (\complex^3)$ under a certain discrete orbifold action. Roughly speaking the main idea is to let a certain discrete group $\Gamma$ act on $\complex^3$ with an appropriate lift to the moduli space $\scrM^{\rm inst}_{n , 0 ; r} (\complex^3)$ in such a way that the isotypical decomposition of the space $W$ is related with the vector spaces in the flag  which characterizes the parabolic structure as in (\ref{Wdec2}) below. A similar construction has appeared in \cite{feigin,finkel,negut} in the case of four dimensional gauge theories and in \cite{biswas} in general.

Consider the group $\Gamma = \zed_r$ which  acts on the target space coordinates as
\begin{equation}
(z_1 , \, z_2 , \, z_3 ) \longrightarrow (z_1 , \, z_2 , \, \omega \, z_3) \ ,
\end{equation}
where $r \in \zed$ and $\omega = \e^{\frac{2 \pi \ii}{r}}$. In particular note that $\cD$ is invariant under this action. We let this group act also on $W = \langle w_1 , \dots , w_r \rangle$ via $\gamma (w_l) = \e^{\frac{2 \pi \ii l}{r}} \, w_l$. As a consequence $W$ decomposes into isotypical components
\begin{equation}
W = \bigoplus_{s \in \hat{\Gamma}} W_{s} \otimes \rho^{\vee}_s \ ,
\end{equation}
where we sum over all the irreducible representations of $\hat{\Gamma}$. Since the defect is full, each factor has $\dim W_{s} = 1$. It will be useful in the following to use this decomposition in the framing condition. This is done via the identification
\begin{equation} \label{Wdec2}
W^{(i)} = \bigoplus_{a=0}^{i-1} W_a  \ .
\end{equation}
The framing condition can be now equivalently expressed in terms of the vector spaces $W_a$, $a=0, \dots , r-1$.
Consider now the covering map 
\begin{eqnarray}
\sigma : \PP^1_{z_1} \times \PP^1_{z_2} \times \PP^1_{z_3} &\longrightarrow&  \PP^1_{z_1} \times \PP^1_{z_2} \times \PP^1_{z_3} \cr
(z_1 , \, z_2 , \, z_3 ) &\longrightarrow& (z_1 , \, z_2 , \, z_3^r ) \ .
\end{eqnarray}
Following  \cite{feigin,finkel} this map can be used to construct an isomorphism $\scrP_{\mbf d} \longrightarrow \scrM^{\rm inst}_{\mbf d , 0 ; r} (\complex^3)^{\Gamma}$, where by $\scrM^{\rm inst}_{\mbf d , 0 ; r} (\complex^3)$ we label the connected component of $\scrM^{\rm inst}_{d , 0 ; r} (\complex^3)$ identified by the decomposition $d = d_0 + \cdots + d_{r-1}$ of the instanton configuration. This isomorphism associates a parabolic sheaf $\cF_{\bullet}$ to a $\Gamma$--equivariant sheaf $\tilde{\cF}$. Note that we already have an obvious morphism of $\scrP_{\mbf d}$ into $\scrM^{\rm inst}_{n , 0 ; r} (\complex^3)$ by forgetting the flag.

To be concrete, consider a parabolic sheaf $\cF_\bullet$. From $\cF_{\bullet}$ we construct the following $\Gamma$-invariant torsion free sheaf on $\PP^1 \times \PP^1 \times \PP^1$
\begin{equation} \label{Gammainv1}
\tilde{\cF} = \sigma^* \cF_{-r+1} + \sigma^* \cF_{-r+2} (-\cD) + \cdots + \sigma^* \cF_{0} (-(r-1) \cD) \ ,
\end{equation}
where the sum is not a direct sum but, as stressed in \cite{negut,feigin}, refers to the convex hull of those sheaves as subsheaves of $\sigma^* \cF_0$. For example, for every open neighbor $\mathcal{U}$, one lists down all the sections for each summand, and then constructs a globally defined sheaf out of them. In particular $\tilde{\cF}$ has rank $r$. Note that the divisor $\cD$ is described by $\{ z_3 = 0 \}$ in local coordinates, and since the action of $\sigma$ on $z_3$ gives $z_3^r$, the inverse image sheaves (which for every open subset $\mathcal{U}$ are given by $\sigma^* \cF_l (\mathcal{U}) = \cF_l (\sigma (\mathcal{U}))$) are by definition $\Gamma$-invariant. The sheaf $\tilde{\cF}$ is constructed by making an invariant sheaf out of each subsheaf in the flag and tensoring it with line bundles of the form $\cO (-k \cD)$ which are associated with characters of $\hat{\Gamma}$ via the action of $\Gamma$ on their sections. The average over the group characters gives a $\Gamma$-invariant sheaf which contains all the information of the original flag. In a sense these line bundles play a role analog to the tautological bundles in the McKay correspondence, as reviewed in \cite{Cirafici:2012qc}.

The sheaf $\tilde{\cF}$ has 
\begin{eqnarray}
c_1 (\tilde{\cF}) &=& - \sum_{i=0}^{r-1} i \ [\cD] \ , \cr
c_3 (\tilde{\cF})&=& d_0 + \cdots + d_{r-1} \ ,
\end{eqnarray}
and vanishing $c_2$; at infinity, it is locally free and framed
\begin{equation}
\tilde{\cF} \vert_{\cD_{\infty}} = \cO_{\cD_{\infty}} \oplus \cO_{\cD_{\infty}} (-\cD) \oplus \cdots \oplus \cO_{\cD_{\infty}} (- (r-1) \cD) \ .
\end{equation}
Therefore given a parabolic sheaf we can construct a $\Gamma$-invariant torsion free sheaf. This is however not quite what we wanted, since $\tilde{\cF}$ has a non vanishing first Chern class, and is therefore not in our moduli space $\scrM^{\rm inst}_{\mbf d , 0 ; r} (\complex^3)$. This can be easily solved by tensoring with a line bundle to cancel the unwanted Chern class (which is of course an isomorphism). A more elegant and direct way is to modify the definition (\ref{Gammainv1}) as suggested in \cite{finkel}
\begin{equation} \label{Gammainv2}
\tilde{\cF} = \sigma^* \cF_{-r+1} + \sigma^* \cF_{-r+2} (-(\cD - \PP^1_{z_1} \times \PP^1_{z_2} \times \infty_{z_3})) + \cdots + \sigma^* \cF_{0} (-(r-1)( \cD- \PP^1_{z_1} \times \PP^1_{z_2} \times \infty_{z_3})) \ .
\end{equation}
The divisor $\PP^1_{z_1} \times \PP^1_{z_2} \times \infty_{z_3} $ is cohomologous to $\cD$ and is used just to cancel its Chern classes. This ensures \cite{finkel} that the map is indeed in $\scrM^{\rm inst}_{n , 0 ; r} (\complex^3)$. Moreover, since $\tilde{\cF}$ is $\Gamma$-invariant, the map is really into $\scrM^{\rm inst}_{n , 0 ; r} (\complex^3)^{\Gamma}$ as we wanted.

On the other hand, given a $\Gamma$-invariant sheaf $\tilde{\cF} \in \scrM^{\rm inst}_{n , 0 ; r} (\complex^3)^{\Gamma}$, we can construct a flag by pasting together the $\Gamma$-isotypical subsheaves. We obtain the flag $\cF_0 (-\cD) \subset \cF_{-r+1} \subset \cdots \subset \cF_0$, having defined
\begin{equation} \label{flagfromGamma}
\cF_{k} = \sigma_* \left( \tilde{\cF} \otimes \cO_X (k \, \cD) \right)^{\Gamma} \ .
\end{equation}
Note that indeed $\ch_1 (\cF_{k}) = k [\cD]$. The line bundle $\cO_X (k \, \cD)$ has a natural orbifold structure. Recall that the divisor $\cD$ is described by $ \{ z_3=0 \}$. Therefore sections of $\cO_X (k \cD)$ are rational functions of the coordinates with a zero (or a pole) at $z_3=0$ of order $k$. Equivalently taking the $\Gamma$-invariant part is going to select the $\omega^{k}$-isotypic component of $\tilde{\cF}$; this is precisely how we recover the flag. In summary we have
\begin{equation}
\scrM^{\rm inst}_{n , 0 ; r} (\complex^3)^{\Gamma} = \bigcup_{|\mbf d| = n} \ \scrP_{\mbf d} \ .
\end{equation}

\subsection{More general divisor defects}

Finally we will relax the condition that the divisor defect is full and consider the most general situation. In words this can be simply done by modeling the defect on a  parabolic group, as done in Section \ref{divisor}, and requiring the flag structure to be the one stabilized by the parabolic group. To simplify the discussion we will firstly outline a ``point-like"  version of the argument and afterwards make our reasoning local in terms of sheaves.

The most general divisor defect is classified by a $M$-tuple of integers $(r_0 , \dots , r_{M-1})$ such that $r_0 + \cdots + r_{M-1} = r$. The factors $r_a$ are the multiplicities of the elements $\alpha_i$ in the reference connection which specifies the divisor defect; the defect will be called of type $\{ r_a \}$. Consider a collection of vector spaces $W_a$ such that $\dim_{\complex} W_a = r_a$ for $a = 0 , \dots , M-1$. Define now the vector spaces $W^{(i)} = \bigoplus_{a  = 0}^{i-1} W_a $ with $i=1,\dots,M$. In particular $\dim_{\complex} W^{(i)}  = \sum_{a = 0}^{i-1} \dim_{\complex} \, W_a  = r_ 0 + \dots + r_{i-1}$. Out of these data we construct the flag
\begin{equation}
W^{(1)} \subset W^{(2)} \subset \cdots \subset W^{(M)} \ .
\end{equation}
This flag correspond to a divisor operator of type $\{ r_a \}$.

Now let us make this construction local, and define the moduli space $\scrP_{\mbf d}^{\{ r_a \}}$ of generalized instantons in the presence of a divisor defect of type $\{ r_a \}$. This moduli space consists of flags of torsion free sheaves of rank $r$ on $\PP^1 \times \PP^1 \times \PP^1$
\begin{equation}
\cF_{0} (- \cD) \subset \cF_{-M+1} \subset \cdots \subset \cF_{-1} \subset \cF_0  \ ,
\end{equation}
such that
\begin{itemize}
\item[(1)] \textit{Framing}. The sheaves in the flag are locally free on $\cD_{\infty}$, together with a framing isomorphism
\begin{equation} \label{framing}
\xymatrix{
 \cF_{0} (- \cD) \vert_{\cD_{\infty}} \ar[r] \ar[d]^\simeqrot & \cF_{-M+1} \vert_{\cD_{\infty}}  \ar[r] \ar[d]^\simeqrot & \cdots \ar[r] \ar[d]^\simeqrot &   \cF_0 \vert_{\cD_{\infty}}  \ar[d]^\simeqrot \\
\cO^{\oplus r}_{\cD_{\infty}} (-\cD) \ar[r] & W^{(1)} \otimes \cO_{\cD_{\infty}} \oplus \cO^{\oplus r-r_0}_{\cD_{\infty}} (-\cD) \ar[r] & \cdots \ar[r]  & W^{(M)} \otimes \cO_{\cD_{\infty}}
}
\end{equation}
where $W^{(i)} $ are the $(r_ 0 + \dots + r_{i-1})$-dimensional vector spaces corresponding to the flag
\begin{equation} \label{flag}
W^{(1)} \subset W^{(2)} \subset \cdots \subset W^{(M)} \ ,
\end{equation}
as outlined above. When this is the standard complete flag we recover the full divisor defect.
\item[(2)] \textit{Chern character}. The framing condition implies $\ch_1 (\cF_k) = - (r - \sum_{a  = 0}^{M-1+k} \, r_a) \, [\cD]$ where $[\cD]$ denotes the fundamental class. We furthermore require $c_2 (\cF_i) = 0$ and $c_3 (\cF_i) = - d_i$. In other words the Chern classes are specified by the degree of the parabolic sheaf  $\mbf d = (d_0 , \cdots , d_{M-1})$ and by the labels $\{ r_a \}$ via the framing condition.
\end{itemize}

The map between $\scrP_{\mbf d}$ and a fixed point subset of $ \scrM^{\rm inst}_{n , 0 ; r} (\complex^3)$ generalizes to $\scrP_{\mbf d}^{ \{ r_a \} }$ upon choosing an appropriate $\Gamma$ action. Note that the only difference between the case of a full divisor defect and the general case is in the Lie algebra structure, and therefore we only need to generalize the action of $\Gamma$ on the vector spaces $W^{(i)}$. Consider the group $\Gamma = \zed_M$ which  acts on the target space coordinates as
\begin{equation}
(z_1 , \, z_2 , \, z_3 ) \longrightarrow (z_1 , \, z_2 , \, \omega \, z_3) \ ,
\end{equation}
where $M \in \zed$ and $\omega = \e^{\frac{2 \pi \ii}{M}}$. This group acts also on $W$ and as a consequence $W$ decomposes into isotypical components
\begin{equation}
W = \bigoplus_{a \in \hat{\Gamma}} W_{a} \otimes \rho^{\vee}_a \ .
\end{equation}
Here $\dim_{\complex} W_a = r_a$ and $\Gamma$ acts as $\gamma (w_l) = \e^{\frac{2 \pi \ii a}{M}} \, w_l$ for $w_l$ any generator of $W_a$. This action is essentially the same as in the full case, but now the $W_a$ are not necessarily unidimensional but take into account the multiplicities $\{ r_a \}$. Exactly has we have explained before, the parabolic structure is encoded in the flag of the vector spaces  $W^{(i)} = \bigoplus_{a  = 0}^{i-1} W_a $. The previous results are recovered for $M=r$. The covering map associated with $\Gamma$
\begin{eqnarray}
\sigma : \PP^1_{z_1} \times \PP^1_{z_2} \times \PP^1_{z_3} &\longrightarrow&  \PP^1_{z_1} \times \PP^1_{z_2} \times \PP^1_{z_3}  \cr
(z_1 , \, z_2 , \, z_3 ) &\longrightarrow& (z_1 , \, z_2 , \, z_3^M ) \ ,
\end{eqnarray}
can now be used to construct explicitly the identification
\begin{equation}
\scrM^{\rm inst}_{n , 0 ; r} (\complex^3)^{\Gamma} = \bigcup_{|\mbf d| = n} \ \scrP_{\mbf d}^{ \{ r_a \} } \ ,
\end{equation}
by a generalization of the previous arguments. Given a parabolic sheaf $\cF_\bullet \in  \scrP_{\mbf d}^{ \{ r_a \}}$ we construct the $\Gamma$-invariant torsion free sheaf
\begin{equation}
\tilde{\cF} = \sigma^* \, \cF_{-M+1} + \sigma^* \cF_{-M+2} (-\cD) + \cdots + \sigma^* \cF_0 ( - (M - 1) \cD) \ .
\end{equation}
At infinity $\tilde{\cF}$ is locally free and framed
\begin{equation}
\tilde{\cF} \vert_{\cD_{\infty}} = \cO^{\oplus \, r_0} \oplus \cO^{\oplus \, r_1} (- \cD) \oplus \cdots \oplus \cO^{\oplus \, r_{M-1}} (- (M-1) \cD) \ ,
\end{equation}
and has
\begin{eqnarray}
c_1 (\tilde{\cF}) &=& - \sum_{a=0}^{M-1} \ a \, r_a \, [\cD] \ ,\\
c_3 (\tilde{\cF}) &=& d_0 + \cdots + d_{M-1} \ ,
\end{eqnarray}
with vanishing $c_2$. Conversely given a $\Gamma$-invariant sheaf $\tilde{\cF}$ we can construct a flag from its isotypical decomposition by using (\ref{flagfromGamma}) as explained before. Note that these maps depend sensitively on the precise action of the orbifold group $\Gamma$. We will use and expand upon this identification in the next section to explain how to compute Donaldson-Thomas type of invariants in the presence of a divisor defect.

\section{Defects and instanton quivers} \label{defects}

In the previous Section we have argued that the problem of Donaldson-Thomas theory in the presence of a divisor operator, in the simple case of the affine space, can be reduced to counting $\Gamma$-equivariant instanton configurations on $\complex^3$ where $\Gamma$ is an appropriate orbifold action determined by the defect. The task of studying the $\Gamma$-fixed point set of $\scrM^{\rm inst}_{n , 0 ; r} (\complex^3)$ is greatly simplified by the knowledge of a local model. In particular the orbifold group and the toric group commute, and one can use virtual localization by considering the set of all torus fixed points of  $\scrM^{\rm inst}_{n , 0 ; r} (\complex^3)$ which are also invariant under the action of $\Gamma$. A formalism to deal with these situations was developed in \cite{Cirafici:2010bd} albeit in a rather different context. This formalism is based on the introduction of a certain quiver quantum mechanics; we will used it to compute the equivariant integrals over the instanton moduli spaces in the presence of the divisor operator.

We should however stress that the problem we are discussing here is rather different from \cite{Cirafici:2010bd}. In particular we do \textit{not} claim that our $\Gamma$-equivariant sheaves on $\complex^3$ are in any sense related to sheaves on a resolution of $\complex^3 / \Gamma$ in any chamber. Indeed such a claim would be wrong: a necessary condition in the construction of \cite{Cirafici:2010bd} is that the $\Gamma$ action on $\complex^3$ has trivial determinant. Technically this ensures that ADHM algebra is Koszul. This fact was used repeatedly in \cite{Cirafici:2010bd} to establish the relation between equivariant sheaves on $\complex^3$ and BPS states in the noncommutative resolution chamber of resolved Calabi-Yau singularities. This is not true in our case. However the fact that this relation does not hold, does not prevent us to consider the $\Gamma$-fixed component of the set of $\torus^3 \times U(1)^r$-fixed sheaves on $\complex^3$, which is always a legitimate procedure. We can indeed think of the formalism of  \cite{Cirafici:2010bd} as an abstract tool to compute equivariant integrals associated to a quiver quantum mechanics, regardless of whether the underlying quiver as a geometric origin of not. 

 In our case the orbifold group $\Gamma$ acts on the geometrical coordinates as
\begin{equation}
(z_1 \, ,  z_2 \, , z_3 ) \longrightarrow (z_1 \, , z_2 \, , \omega \, z_3) \ ,
\end{equation}
where $\omega = \e^{\frac{2 \pi \ii}{M} }$. Similarly we let it act on the Cartan subalgebra generators of $U(r)$ as
\begin{equation} \label{orboncartan}
(a_1 , \cdots , a_r) \longrightarrow \left(  \underbrace{ a_1 , \cdots  , a_{r_0} }_{r_0} , \underbrace{ \omega \, a_{r_0+1} , \cdots , \omega \, a_{r_0 + r_1} }_{r_1} , \cdots , \underbrace{ \omega^{M-1} \, a_{r-r_{M-1}} , \cdots , \omega^{M-1} \, a_r }_{r_{M-1}}\right)  \ ,
\end{equation}
where the pattern is determined by the choice of the divisor operator, since it breaks the gauge group to the parabolic subgroup determined by the parameter $\alpha$. The breaking is parametrized by a set of integers $\{ r_a \}$ with $a = 0 , \dots , M-1$ and $r = r_0 + \cdots + r_{M-1}$. We will often use the convenient notation $a_{I,s}$ to collect all the Cartan generators upon which $\Gamma$ acts as $\omega^I$, with $s=1,\dots,r_I$. Note that we are assuming that the Cartan generators are (eventually after a gauge transformation) ordered. This choice of action is directly related to the decomposition (\ref{Wdec2}) which enters in the description of the moduli space of parabolic torsion free sheaves. Indeed this identification is manifest if we take $\dim_{\complex} W_a = r_a$ or equivalently $W_a$ has generators $\e^{\ii a_{a,s}}$ as a $U(1)^{r_a}$ module. This is just another way to see explicitly the identification between parabolic sheaves determined by the flag (\ref{flag}) and orbifold sheaves determined by the action (\ref{orboncartan}).

The orbifold action lifts to the instanton moduli space. To describe this lift it is convenient to use the ADHM formalism introduced previously and give directly the orbifold action on the generalized ADHM data. As a consequence we will be able to describe the relevant instanton configurations in terms of a certain quiver, obtained from the ADHM quiver by decomposing the maps and the vector spaces according to the orbifold characters \cite{Cirafici:2010bd}. We decompose the vector spaces $V$ and $W$ as follows
\begin{equation}
V = \bigoplus_{a\in\Gammaw}\, V_a \otimes
\rho_a^{\vee} \ , \qquad W = \bigoplus_{a\in\Gammaw}\, W_a \otimes \rho_a^{\vee} \ ,
\end{equation}
such that $\dim_{\complex} V_a = n_a$ and $\dim_{\complex} W_a = r_a$ and $\{ {\rho_a} \}_{a\in\widehat{\Gamma}}$ is the set of irreducible
representations; we denote the trivial representation by $\rho_0$. 
In particular now the bosonic field content of the matrix quantum mechanics is made up by equivariant maps
\begin{eqnarray}
(B_1 , B_2 , B_3 , \varphi) \in\Hom_{\Gamma} (V,V) \qquad \mbox{and} \qquad
I\in\Hom_{\Gamma} (W,V) \ .
\end{eqnarray}
More explicitly the only non-vanishing components are
\begin{eqnarray}
B_{1,2}^{a} \, &:& \, V_{a} \ \longrightarrow \ V_{a} \ , \nonumber \\[4pt]
B_{3}^{a} \, &:& \, V_{a} \ \longrightarrow \ V_{a+1} \ , \nonumber \\[4pt]
\varphi^{a} \, & : & \, V_{a} \ \longrightarrow \ V_{a+1} \ , \nonumber \\[4pt]
I^{a} \, & : & \, W_{a} \ \longrightarrow \ V_{a} \ .
\end{eqnarray}
These maps are associated to the following BRST transformations
\begin{equation}
\begin{array}{rllrl}
\cQ \, B_\a^{a} &=& \psi_\a^r & \quad \mbox{and} \qquad \cQ \, \psi_\a^{a}& = \ [\phi , B_\a^{a}] -
\epsilon_\a\,
B_\a^{r} \ , \\[4pt] \cQ \, \varphi^{r}& =& \xi^{r} & \quad \mbox{and} \qquad
\cQ \, \xi^{a} &= \ [\phi ,\varphi^{a}] - (\epsilon_1 + \epsilon_2 + \epsilon_3 ) \varphi^{a}  \ , \\[4pt]
\cQ \, I^{a} &=& \varrho^{a} & \quad \mbox{and} \qquad \cQ \, \varrho^{a} & = \
\phi \,I^{a} - I^{a} \,\mbf a^{a} \ ,
\end{array}
\end{equation}
where in the vector $\mbf a^{a}$ we have collected all the Higgs field eigenvalues $a_l$ associated with the irreducible representation $\rho_a$. Following the standard formalism of topological field theories, one associates to these maps two Fermi multiplets containing the anti-ghosts and the auxiliary fields, and an extra gauge multiplet to close the BRST algebra \cite{Cirafici:2010bd}. Then one proceed to construct a topological invariant action which localizes onto the critical points of the BRST operator. 
These datas can be conveniently summarized in the generalized ADHM quiver 
\begin{equation} \label{ADHMgen}
\vspace{4pt}
\xymatrix@C=20mm{
 \cdots  \ar@//[r]^{B^{a-2}_3}
\ar@{.>}@/_/[r]_{\varphi^{a-2}} 
& \ V_{a-1 }\ \bullet \ \ar@(u,ul)_{B^{a-1}_2} \ar@(u,ur)^{B^{a-1}_1} \ar@//[r]^{B^{a-1}_3}
\ar@{.>}@/_/[r]_{\varphi^{a-1}} 
& \ V_a \ \bullet \ \ar@(u,ul)_{B^a_2} \ar@(u,ur)^{B^a_1} \ar@//[r]^{B^a_3}
\ar@{.>}@/_/[r]_{\varphi^a} 
& \ V_{a+1} \ \bullet \ \ar@(u,ul)_{B^{a+1}_2} \ar@(u,ur)^{B^{a+1}_1} \ar@//[r]^{B^{a+1}_3}
\ar@{.>}@/_/[r]_{\varphi^{a+1}} & \cdots
 \\ \\
&  W_{a-1} \ \bullet \ar@//[uu]^{I^{a-1}} 
& W_a \ \bullet \ar@//[uu]^{I^a} 
& W_{a+1} \ \bullet \ar@//[uu]^{I^{a+1}} 
&
}
\vspace{4pt}
\end{equation}
In particular to this modified quiver one associates an ideal of relations which arises from decomposing the original ADHM equations accordingly to the $\Gamma$--module structure. Recalling that we are interested in the set of minima where the field $\varphi$ is set to zero, the relevant equations are
\begin{equation} \label{surfADHM}
\begin{array}{rllrl}
B_1^a \, B_2^a & - & B_2^a \, B_1^a&=& 0 \, ,\\
 B_1^{a+1} \, B_3^a & - & B_3^a \, B_1^a &=& 0 \, , \\
 B_2^{a+1} \, B_3^a & - & B^a_3 \, B_2^a &=& 0 \,  ,\\ 
& & \left( I^{a+1} \right)^{\dagger} \ \varphi^a &=& 0 \, . 
\end{array}
\end{equation}
These equations generate the ideal of relations in
the instanton quiver path algebra $\sfA_\Gamma$. Their $\Gamma$-equivariant decomposition cuts out a certain subvariety ${\sf Rep}_{\Gamma} (\mbf n , \mbf r;B)$ from the framed quiver representation space
\begin{equation}
{\sf Rep}_{\Gamma} (\mbf n , \mbf r) =  \Hom_{\Gamma} (V , Q \otimes V) \ 
 \oplus \ \Hom_{\Gamma} (V , \mbox{$\bigwedge^3$} Q \otimes V) \ 
\oplus \ \Hom_{\Gamma} (W , V) \ ,
\end{equation}
The BPS moduli space in the presence of a divisor defect is then formally defined as the quotient stack
\begin{equation}
\scrM_{\Gamma} (\mbf n ,\mbf r) = \Big[ {\sf Rep}_{\Gamma} (\mbf n , \mbf r;B) \,
\Big/ \, \prod_{a\in\widehat \Gamma}\, GL(n_a,\IC) \Big]
\end{equation}
by the gauge group which acts as basis change automorphisms of the
$\Gamma$-module $V$. As in \cite{Cirafici:2010bd}, we will think of this stack as a moduli space of stable framed representations, where every object in the category of quiver representations with relations is $0$-semistable.

We define our Donaldson-Thomas invariants in the presence of a divisor operator as the equivariant volumes of these instanton moduli spaces, computed via virtual localization. In doing so we are making explicit use of the fact that the relevant moduli space is a $\Gamma$-fixed component of the moduli space of torsion free sheaves, whose toric fixed points are isolated and given by ideal sheaves. Therefore all the relevant machinery of virtual localization can be applied directly to the case at hand. Of course the problem of giving a rigorous definition of these invariants on a generic variety is still quite open; for the time being we will limit ourselves to the case of $\complex^3$. Note that it is conceivable that the approach we are following generalizes to any toric Calabi-Yau, since one can imagine to carry on the construction of Section \ref{parabolic} on toric invariant patches and find appropriate gluing rules to construct sheaves on a generic toric variety with a defect. We plan to return to this problem (and its lower dimensional version on a toric surface) as well as to the task of giving rigorous definitions on any Calabi-Yau, in the future. On $\complex^3$ we have
\begin{equation}
\DT_{n , r}^{\cD} \left( \complex^3 \ \vert \ \epsilon_1 , \epsilon_2 , \epsilon_3 , a_l \right) = \int_{[ \scrM_{\Gamma} (\mbf n ,\mbf r) ]^{\rm vir}} \ 1 \ ,
\end{equation}
where the right hand side takes values in the polynomial ring $\mathbb{Q} [\epsilon_1 , \epsilon_2 , \epsilon_3 , a_1 , \dots , a_r]$ as in Section \ref{instanton}. We do not impose the condition $\epsilon_1 + \epsilon_2 + \epsilon_3 = 0 $,  and therefore the orbifold group $\Gamma$ is a discrete subgroup of the toric group and we can simply restrict our attention to toric fixed points which are simultaneously $\Gamma$-fixed. In particular the invariants are only  equivariant and depend explicitly on the parameters of the toric action, as was expected from the four dimensional case. They are essentially noncommutative Donaldson-Thomas type of invariants associated with the quiver moduli space $\scrM_{\Gamma} (\mbf n ,\mbf r)$. Note however that, contrary to the standard noncommutative Donaldson-Thomas invariants of \cite{szendroi}, they depend explicitly on the vector $\mbf r$. From this perspective they are more similar to the invariants $\NDT_{\mu} (\mbf n , \mbf r)$ studied in \cite{Cirafici:2010bd,Cirafici:2011cd}, with the important difference that the $\mbf r$ dependence is genuine and cannot be encoded in a multiplicative factor. Physically this is clear; the $\mbf r$ dependence in the $\NDT_{\mu} (\mbf n , \mbf r)$ invariants label a gauge theory superselection sector, characterized by different boundary conditions for the gauge field at infinity, but the dynamics in each sector is essentially the same. Here different vectors $\mbf r$ label physically inequivalent defects, characterized by different parabolic subgroups. 

A local model of the virtual tangent space $T_{[\cE_{\vec{\pi}}]}^{\mathrm{vir}}  \scrM_{\Gamma} (\mbf n ,\mbf r)$ at a fixed point $\cE_{\vec{\pi}}$ labelled by $\vec{\pi}$, is given by the instanton deformation complex
\begin{equation} \label{equivdefcomplex}
\xymatrix{
  \Hom_{\Gamma}  (V_{\vec\pi} , V_{\vec\pi})
   \quad\ar[r]^{\!\!\!\!\!\!\!\!\!\!\!\!\!\!\!\!\sigma} &\quad
   {\begin{matrix} \Hom_{\Gamma} (V_{\vec\pi} , V_{\vec\pi} \otimes Q )
   \\ \oplus \\
   \Hom_{\Gamma} (W_{\vec\pi} , V_{\vec\pi}) \\ \oplus  \\ \Hom_{\Gamma} (V_{\vec\pi} ,
   V_{\vec\pi}  \otimes \bigwedge^3 
   Q) \end{matrix}}\quad \ar[r]^{\tau} & \quad
   {\begin{matrix} \Hom_{\Gamma} (V_{\vec\pi} , V_{\vec\pi}  \otimes \bigwedge^2
       Q) \\ \oplus \\ 
       \Hom_{\Gamma} (V_{\vec\pi},W_{\vec\pi} \otimes \bigwedge^3 Q)
   \end{matrix}}
} \ .
\end{equation}
Here the map $\tau$ follows from the linearization of the equations (\ref{surfADHM}) while the map $\sigma$ is an infinitesimal gauge transformation. We have introduced the orbifold regular representation $Q = \rho_0 \oplus \rho_0 \oplus \rho_1$. The Euler class of the virtual tangent space can be computed via the equivariant index of this complex. Neglecting the $\Gamma$-action, the two vector spaces $V$ and $W$ can
be decomposed at a
fixed point $\vec\pi=(\pi_1,\dots,\pi_r)$ of the $U(1)^r \times \torus^3$ action on the
instanton moduli space, as~\cite{Cirafici:2008sn}
\begin{eqnarray}
V_{\vec\pi} = \sum_{l=1}^r \,e_l~ \sum_{(n_1,n_2,n_3)\in \pi_l}\,
t_1^{n_1-1} \,t_2^{n_2-1}\,t_3^{n_3-1} \qquad \mbox{and} \qquad W_{\vec\pi} =
\sum_{l=1}^r\,e_l \ ,
\label{decompos}
\end{eqnarray}
where $e_l=\e^{\ii a_l}$ with $a_l$ the Higgs field vacuum expectation values for $l=1,\dots,r$. The orbifold group $\Gamma$ acts on each of the module generators in the above decomposition with a weight which is determined by the position of the box in the plane partition, as labeled by the three integers $(n_1,n_2,n_3)\in \pi_l$. As a consequence each box will be associated to a character of the orbifold group. In other words $\Gamma$-fixed torus invariant points are labeled by vectors of colored plane partitions, the coloring being associated with distinct characters of the orbifold group.

This decomposition is however inconvenient for certain purposes and it is sometime better to use a different one which disentangles spacetime variables from the Cartan generators. Fixed points of the toric action are $r$--tuples of plane partitions, each one associated with a generator of the Cartan subalgebra of $U(r)$. Since $\Gamma$ acts as (\ref{orboncartan}) on the Cartan parameters, the action on the plane partition is ``offset". This is clear from the decomposition (\ref{decompos}). To keep track of this offset we will introduce a ``defect function" ${\tt d} : \{ 1 , \dots , r \} \longrightarrow \hat{\Gamma}$ which to a sector labelled by $l =1 , \dots , r$ and corresponding to a module $E_{I,s}$ for any $s$, associates the weight of the corresponding representation of $\Gamma$. In other words ${\tt d} (l) = I$ if the module $E_l$ is spanned by $a_{I,s}$ for $s=1 , \dots , r_I$. Then
\begin{equation}
V_{\vec\pi} =\bigoplus_{l=1}^r ~ \bigoplus_{a\in\Gammaw} \, \big( E_l \otimes \rho_{{\tt d} (l)}^{\vee} \big) \otimes \left( P_{l,a} \otimes \rho_a^{\vee} \right) =\bigoplus_{l=1}^r~ \bigoplus_{a\in\Gammaw} \, \big( E_l \otimes P_{l,a} \big) \otimes \rho_{a+{\tt d} (l)}^{\vee} \ .
\label{VpiGammadecomp}\end{equation}
Here $P_{l,a}$ is a module which corresponds to the
$\Gamma$-module decomposition of the sum  $H^0 (\cO_{\cI_{\pi_l}}) =  \sum_{(n_1,n_2,n_3)\in \pi_l}\, t_1^{n_1-1} \,t_2^{n_2-1}\,t_3^{n_3-1} $. Recall that each fixed point is characterized by a vector of partitions $\vec \pi$. Each entry in this vector can be decomposed according to the $\Gamma$-action, taking further into account the transformation properties of the Higgs field vacuum expectation values $e_l$. In our decomposition (\ref{VpiGammadecomp}) we have factorized this contribution explicitly so that now $\dim_\complex P_{l,a}$ is the number of boxes in the plane partition at position $l$ of the fixed point vector $\vec \pi = (\pi_1 , \dots , \pi_r)$ which transform in the representation $\rho_r^{\vee}$, a number which we will call $|\pi_{l,a}|$. Similarly one can write
\begin{equation}
W_{\vec\pi} = \bigoplus_{l=1}^r \, E_l \otimes \rho_{{\tt d} (l)}^{\vee} \ .
\end{equation}

Given this formalism we can now compute the character at a fixed point
\begin{equation} \label{orbcharacter}
\ch_{\torus^3 \times U(1)^r} \left(  T_{\vec{\pi}}^{\mathrm{vir}}   \scrM_{\Gamma} (\mbf n ,\mbf r)   \right)
= \left( W_{\vec\pi}^\vee \otimes V_{\vec\pi} -
\frac{ {V}_{\vec\pi}^\vee \otimes W_{\vec\pi} }{t_1 \, t_2 \, t_3} + \frac{ (1-t_1)\, (1-t_2)\,
(1-t_3) }{t_1 \, t_2 \, t_3} ~ {V}^\vee_{\vec\pi} \otimes V_{\vec\pi} \right)^{\Gamma} \ .
\end{equation}
Let us consider each term separately. First of all
\begin{eqnarray}
\left( W_{\vec\pi}^{\vee} \otimes V_{\vec\pi} \right)^{\Gamma} &=& \bigoplus_{l,l'} \ \bigoplus_{a \in \hat{\Gamma}} \ E_l \otimes P_{l,a} \otimes E^{\vee}_{l'} \otimes \left( \rho_{a+ {\tt d} (l)}^{\vee} \otimes \rho_{{\tt d} (l')} \right)^{\Gamma} \cr
&=&  \bigoplus_{l,l'} \ \bigoplus_{a \in \hat{\Gamma}} \ E_l \otimes P_{l,a} \otimes E^{\vee}_{l'} \ \delta \left( a + {\tt d} (l) = {\tt d} (l') \mod M \right) \ .
\end{eqnarray}
Here we have used the fact that the invariant part of a tensor product between representations is a Kronecker delta
\begin{equation}
\left( \rho^{\vee}_a \otimes \rho_b \right)^{\Gamma} = \delta (a=b \mod M) \ ,
\end{equation}
where we write the constraint explicitly to make the formulas more readable. 
Similarly
\begin{equation}
- \left(\frac{ V_{\vec\pi}^{\vee} \otimes W_{\vec\pi}}{t_1 t_2 t_3} \right)^{\Gamma} = - \frac{1}{t_1 t_2 t_3 }  \bigoplus_{l,l'=1}^r \bigoplus_{a \in \hat{\Gamma}} \, E_l^{\vee} \otimes P^{\vee}_{l,a} \otimes E_{l'}  \ \delta \left( a + {\tt d} (l) = {\tt d} (l') +1 \mod M \right) \ .
\end{equation}
We have used the fact that the weights $t_{\alpha}$ are regarded as $\Gamma$-modules; in particular $t_3 \longrightarrow \rho_1$ corresponds to the action of $\Gamma$ on $z_3$ by multiplication with $\omega$. The remaining terms are
\begin{eqnarray}
&& \hspace{-1.5cm} - \left( \left( 1 - t_1^{-1} \right) \left( 1 - t_2^{-1} \right) V_{\vec\pi}^{\vee} \otimes V_{\vec\pi} \right)^{\Gamma} = -  \left( 1 - t_1^{-1} \right) \left( 1 - t_2^{-1} \right) 
\\[4pt] && \times \,  \nonumber
\left( \bigoplus_{l,l'}^{r} \bigoplus_{a,b \in \hat{\Gamma}} E^{\vee}_l \otimes P_{l,a}^{\vee} \otimes E_{l'} \otimes P_{l',b} \right)  \, \delta \left( a+{\tt d} (l) = b + {\tt d} (l') \mod M \right) \ , \\[4pt] 
&& \hspace{-1.5cm} \left( \left( 1 - t_1^{-1} \right) \left( 1 - t_2^{-1} \right) \, t_3^{-1} \, V_{\vec\pi}^{\vee} \otimes V_{\vec\pi} \right)^{\Gamma} =   \left( 1 - t_1^{-1} \right) \left( 1 - t_2^{-1} \right) \, t_3^{-1} \, 
\\[4pt] && \times \,  \nonumber
\left( \bigoplus_{l,l'}^{r} \bigoplus_{a,b \in \hat{\Gamma}} E^{\vee}_l \otimes P_{l,a}^{\vee} \otimes E_{l'} \otimes P_{l',b} \right)  \times \, \delta \left( a+{\tt d} (l)   = b + 1 + {\tt d} (l') \mod M \right)  \ .
\end{eqnarray}

As explained in Section \ref{instanton} we can now use virtual localization to compute the equivariant volumes of the instanton moduli spaces, or in the language of the quiver quantum mechanics the ratio of the fluctuation determinants around each instanton solution. The invariants are given by
\begin{equation}
\DT_{n , r}^{\cD} \left( \complex^3 \, \vert \, \epsilon_1 , \epsilon_2 , \epsilon_3 , a_l \right) = \int_{ [ \scrM_{\Gamma} (\mbf n ,\mbf r) ]^{\mathrm{vir}} } \ 1 = \sum_{[\cE_{\vec{\pi}}] \in \scrM_{\Gamma} (\mbf n ,\mbf r)^{\torus^3 \times U(1)^r} } \ \frac{1}{\mathrm{eul}_{\torus^3 \times U(1)^r } \left( T_{[\cE_{\vec{\pi}}]}^{\mathrm{vir}}  \scrM_{\Gamma} (\mbf n ,\mbf r)   \right) }  \ .
\end{equation}
As we have already stressed, the invariants depend explicitly on the variables parametrizing the Cartan subalgebra of $\torus^3 \times U(1)^r$, that is we are dealing with an equivariant version of Donaldson-Thomas theory. Sometimes it is convenient to write
\begin{equation}
\DT_{n , r}^{\cD} \left( \complex^3 \, \vert \, \epsilon_1 , \epsilon_2 , \epsilon_3 , a_l \right) = \sum_{\vec{\pi} \, : \, |\vec{\pi}| = n} \ \DT_{\vec{\pi}}^{\cD} \left( \complex^3 \, \vert \, \epsilon_1 , \epsilon_2 , \epsilon_3 , a_l \right) \ ,
\end{equation}
when we want to keep track of the orbifold characters in the combinatorial problem. Each invariant can now be computed explicitly, although the final expression is rather cumbersome. We write it here for completeness:
\begin{eqnarray} 
\DT_{\vec{\pi}}^{\cD} \left( \complex^3 \, \vert \, \epsilon_1 , \epsilon_2 , \epsilon_3 , a_l \right)  \\[12pt] && \hspace{-3cm}
= \prod_{l,l'}^r
\frac{ \displaystyle \prod_{\stackrel{(n_1 , n_2 , n_3) \in \pi_l}{{\tt d} (l)-{\tt d} (l')+n_3-1=0 \mod M}}  \left(-\ii a_l+\ii a_l' - \ii \epsilon_1 n_1 - \ii \epsilon_2 n_2 - \ii \epsilon_3 n_3 \right)}{\displaystyle \prod_{ \stackrel{(n_1 , n_2 , n_3) \in \pi_l}{{\tt d} (l)-{\tt d} (l')+n_3=0 \mod M}}  \left( \ii a_l-\ii a_l'+\ii \epsilon_1 (n_1 -1) + \ii \epsilon_2 (n_2 - 1) + \ii \epsilon_3 (n_3 - 1) \right)}
 \nonumber \\[12pt] && \hspace{-2.5cm} \nonumber
\begin{array}{lcl}
\times & \displaystyle \prod_{\substack{(n_1 , n_2 , n_3) \in \pi_l \, , \, (n'_1 , n'_2 , n'_3) \in \pi_l' \\ -{\tt d} (l)+{\tt d} (l')-n_3+n'_3=0 \mod M}}  & \frac{-\ii a_l + \ii a_l' + \ii \epsilon_1 (n'_1-n_1) + \ii \epsilon_2 (n'_2 - n_2) + \ii \epsilon_3 (n'_3 - n_3)}{-\ii a_l + \ii a_l' + \ii \epsilon_1 (n'_1-n_1-1) + \ii \epsilon_2 (n'_2 - n_2) + \ii \epsilon_3 (n'_3 - n_3)}
\\[12pt]
\times & \displaystyle \prod_{\stackrel{(n_1 , n_2 , n_3) \in \pi_l \, , \, (n'_1 , n'_2 , n'_3) \in \pi_l'}{-{\tt d} (l)+{\tt d} (l')-n_3+n'_3=0 \mod M}}  & \frac{-\ii a_l + \ii a_l' + \ii \epsilon_1 (n'_1-n_1-1) + \ii \epsilon_2 (n'_2 - n_2-1) + \ii \epsilon_3 (n'_3 - n_3)}{-\ii a_l + \ii a_l' + \ii \epsilon_1 (n'_1-n_1) + \ii \epsilon_2 (n'_2 - n_2-1) + \ii \epsilon_3 (n'_3 - n_3)}
\\[12pt]
\times & \displaystyle \prod_{\stackrel{(n_1 , n_2 , n_3) \in \pi_l \, , \, (n'_1 , n'_2 , n'_3) \in \pi_l'}{-{\tt d} (l)+{\tt d} (l')-n_3+n'_3-1=0 \mod M}} &  \frac{-\ii a_l + \ii a_l' + \ii \epsilon_1 (n'_1-n_1-1) + \ii \epsilon_2 (n'_2 - n_2) + \ii \epsilon_3 (n'_3 - n_3-1)}{-\ii a_l + \ii a_l' + \ii \epsilon_1 (n'_1-n_1) + \ii \epsilon_2 (n'_2 - n_2) + \ii \epsilon_3 (n'_3 - n_3-1)}
\\[12pt]
\times & \displaystyle \prod_{\stackrel{(n_1 , n_2 , n_3) \in \pi_l \, , \, (n'_1 , n'_2 , n'_3) \in \pi_l'}{-{\tt d} (l)+{\tt d} (l')-n_3+n'_3-1=0 \mod M}} &  \frac{-\ii a_l + \ii a_l' + \ii \epsilon_1 (n'_1-n_1) + \ii \epsilon_2 (n'_2 - n_2-1) + \ii \epsilon_3 (n'_3 - n_3-1)}{-\ii a_l + \ii a_l' + \ii \epsilon_1 (n'_1-n_1-1) + \ii \epsilon_2 (n'_2 - n_2-1) + \ii \epsilon_3 (n'_3 - n_3-1)}
\end{array} \ .
\end{eqnarray}

This formula can be expanded by summing explicitly on a vector of partitions $\vec{\pi}$. Since the results are not very illuminating we do not include them here.

Finally we can collect all our results and write down, at least formally, the generating function of Donaldson-Thomas invariants in the presence of a divisor defect $\cD$
\begin{eqnarray}
\cZ^{DT}_{(\complex^3 , \cD)} (q \, \vert \, \epsilon_1 , \epsilon_2 , \epsilon_3 , a_l  ) &=& \sum_{\vec{\pi}}  \ \DT_{\vec{\pi}}^{\cD} \left( \complex^3 \, \vert \, \epsilon_1 , \epsilon_2 , \epsilon_3 , a_l \right) \ \prod_{a \in \hat{\Gamma}} \ q_a^{\sum_{l=1}^r \vert \pi_{l,a-{\tt d} (l)} \vert} \cr
&=& \sum_{n} \  \DT_{n , r}^{\cD} \left( \complex^3 \, \vert \, \epsilon_1 , \epsilon_2 , \epsilon_3 , a_l \right) \ q^n \ ,
\end{eqnarray}
where we have introduced the formal parameters $q_a$ to keep track of the instanton numbers associated to each character of the orbifold group, and $q$ for the overall instanton number. We stress that this partition function can be computed explicitly term by term in the variables $q_a$ or $q$.

\section{Defects in higher dimensional theories} \label{defectsH}

It is natural to expect that some of our results hold for other cohomological field theories and in other dimensions as well. These theories have been much studied in the past \cite{Baulieu:1997jx,Hofman:2000yx,Acharya:1997gp,Blau:1997pp,Baulieu:1997nj,Dijkgraaf:2004te}. One starts with a manifold $M_N$ of real dimension $N$ which is endowed with a certain structure. The most interesting case is when this structure is associated with a reduction of the holonomy. Generically to this problem one can associate the generalized instanton equations
\begin{equation}
\lambda \, F^{\mu \nu} = \frac{1}{2} \, T^{\mu \nu \rho \sigma} \, F_{\rho \sigma}  \ ,
\end{equation}
where $\lambda$ is a constant parameter while $T^{\mu \nu \rho \sigma}$ is a certain antisymmetric tensor. Here the indices run from $\mu=1,\dots,N$ and the tensor $T^{\mu \nu \rho \sigma}$ is responsible for reducing the holonomy from $SO(N)$ to a subgroup. For example if the holonomy is reduced to $Spin(7)$ the tensor can be chosen to be
\begin{equation}
T^{\mu \nu \rho \sigma} = \zeta^{T} \ \gamma^{\mu \nu \rho \sigma} \zeta \ .
\end{equation}
Here $\gamma^{\mu \nu \rho \sigma} $ is the totally antisymmetric product of $\gamma$ matrices for the $SO(8)$ spinor representation, while $\zeta$ is the covariantly constant spinor corresponding to the singlet in the decomposition of the chiral spinor representation of $SO(8)$ induced by the reduced holonomy. This choice leads to an interesting theory with one topological BRST charge based on a certain octonionic generalization of the instanton equations \cite{Baulieu:1997jx}.

In this section we will consider briefly only a specific example of another cohomological theory: the eight dimensional theory obtained when the holonomy is reduced to $SU(4)$ by choosing as tensor $T$ the holomorphic $(4,0)$ form $\Omega$. This theory is essentially a eight dimensional version of the theory we have been studying so far and was throughly analyzed in \cite{Baulieu:1997jx}. Given $\Omega$ one can define the operator $*$ on $M_8$ as
\begin{equation}
* \ : \ \Omega^{0,p} (M_8)  \longrightarrow \Omega^{0,4-p} (M_8) \ ,
\end{equation}
via the pairing
\begin{equation}
\langle \alpha , \beta \rangle = \int_{M_8} \ \Omega \wedge \alpha \wedge * \beta \ .
\end{equation}
We denote with $\Omega^{0,2}_{\pm} (M_8)$ the eigenspaces of $*$ and let $P_{\pm}$ be the projection. Consider an holomorphic bundle $\cE$. Then we call a connection $\overline{\partial}_A$ with $F_A^{0,2} = \overline{\partial}_A^2$ holomorphic anti-self-dual if $P_+ F_A^{0,2}=0$. Given an holomorphic anti-self-dual connection the complex of  adjoint valued differential forms
\begin{equation}
\label{8dcomplex}
\xymatrix@1{0 \ar[r] & \Omega^{0,0} ( M_8 , \mathrm{ad}\,  \cE)
\ar[r]^{\hspace{-0.2cm} \overline{\partial}_A} & ~\Omega^{0,1} ( M_8 , \mathrm{ad}\, \cE) 
\ar[r]^{\hspace{0.2cm} P_+ \overline{\partial}_A} & \Omega^{0,2} ( M_8 , \mathrm{ad}\, \cE) \ar[r] & 0 } \ ,
\end{equation}
is elliptic\footnote{The reason why (\ref{defcomplex}) has a $\Omega^{0,3}$ in the complex is that this is dual to a scalar which descends from a component of the gauge field in $\Omega^{0,1}$  via dimensional reduction.}. The cohomological theory is obtained by gauge fixing the topological invariant
\begin{equation}
S_8 = \int_{M_8} \Omega \wedge \Tr (F_A^{0,2} \wedge F^{0,2}_A) \ .
\end{equation}
The theory can be studied with standard cohomological techniques \cite{Baulieu:1997jx}, and it localizes onto the moduli space of holomorphic anti-self-dual connections in the topological sector with $S_8$ fixed. These configurations correspond to generalized instanton solutions. As usual we denote by $\scrM^{\rm inst}_{\ch_i ; r} (M_8)$  the moduli space of instanton solutions modulo complexified gauge transformations and fixed Chern characters $\ch_i$. The intersection theory of this moduli space, albeit yet to be rigorously defined, is expected to give a higher dimensional generalization of Donaldson theory \cite{Baulieu:1997jx}.

We now consider the theory defined with a divisor defect on a divisor $D_6$. The formalism developed in Section \ref{divisor} can be applied almost verbatim. The defect has co-dimension two and we require gauge connections to have a prescribed monodromy $\alpha$ around the position of the defect, associated with a Levi subgroup $L$ of $G$. Equivalently one can deal with $G$-bundles over all of $M_8$, whose structure group is reduced to $L$ along $D_6$. Once the theory is modified by the inclusion of the defect, one can introduce instantons and study the moduli problem associated with the combination $F_A - 2 \pi \alpha \, \delta_{D_6}$. Therefore we are led to the moduli space 
\begin{equation} 
\scrM^{\alpha} (L ; M_8) = \left. \left\{ A \in  \Omega^{(0,1)} \left( M_8 ; \mathrm{ad} \,  \cE  \right)  \, \Big{\vert} \   
P_+ \, (F_A   -  2 \pi \alpha \delta_{D_6})^{(0,2)} = 0 , \text{stable}
\right\} \right/ \mathcal{G}_{D_6} \ ,
\end{equation}
parametrizing holomorphic anti-self-dual connections whose structure group is reduced to $L$ along $D_6$, modulo gauge transformations which take values in $L$ on $D_6$, and an appropriate stability condition. This moduli space is filtered in topological sectors by fixing the Chern characters $\ch_i (\cE)$. Equivalently we can switch to the language of parabolic sheaves and study the moduli spaces $\scrP^{\alpha} (M_8 ; D_6)$ of stable torsion free sheaves with a parabolic structure over $D_6$. In principle one would like to study the intersection theory of these moduli spaces and define enumerative invariants. Unfortunately already the instanton moduli space $\scrM^{\rm inst}_{\ch_i ; r} (M_8)$ is poorly understood and technical difficulties quite challenging. Still, we believe that the geometrical problem associated with the moduli space $\scrP^{\alpha} (M_8 ; D_6)$ is rather interesting and deserves further study.

\section{Discussion} \label{discussion}

\subsection{Divisor defects}

The purpose of this paper is to lay down the foundations of a theory of divisor defects for generic cohomological quantum field theories in higher dimensions. We have discussed in detail one particular case corresponding to Donaldson-Thomas theory of Calabi-Yau threefolds. In plain words we conjecture the existence of a generalization of the Donaldson-Thomas enumerative problem where one replaces the moduli space of stable sheaves with the moduli space of stable sheaves with a certain parabolic structure on a divisor. Regrettably the present work has a somewhat programmatic flavor, as the only explicit (equivariant) computation we could do is in the affine case, and for a very special divisor operator. In this case we can use an alternative description of the moduli space of parabolic sheaves as the fixed point set of the generalized instanton moduli space, with respect to a certain orbifold action determined by the defect. Having established this identification, all the powerful technology of virtual localization can be used to compute explicitly the invariants. In particular one gets as a byproduct rather precise definitions, since all the integrals can be defined via virtual localization and one can use the results already established in the case without the defect. The problem on a generic compact Calabi-Yau seems on the other hand quite intractable. To our knowledge there are very few results in the literature on parabolic sheaves on Calabi-Yau varieties. It would be very interesting to understand if the general discussion outlined in this paper can be given a rigorous treatment.

\subsection{Defects within defects and $k$-categories}

We have repeatedly mentioned the possibility of including other defects supported on the divisor operator. Recall that defects can be thought of as local modifications of the theory, for example by prescribing appropriate boundary conditions on the fields nearby the defect. As a consequence the space of BPS states of the original theory $\cH^{BPS}$ is modified to a new Hilbert space $\cH^{BPS}_\mathsf{D}$ which includes the effects of the defect. The defect itself can be described via an effective field theory, and this effective field theory can support defects on its own. For example, if a theory supports a domain wall, it is effectively governed by two quantum field theories, say theory $\mathsf{A}$ and theory $\mathsf{B}$, interacting only via the domain wall. The original space of states $\cH^{BPS}$ will be further modified, say to $\cH^{BPS}_{\mathsf{D}_\mathsf{A} \, , \mathsf{D}_\mathsf{B}}$. Theories $\mathsf{A}$ and $\mathsf{B}$ can themselves support defects, and so on, leading to an intricate but extremely rich picture. From our point of view, each further layer of defects corresponds to a different enumerative problem; naively a collection of defects within defects $\{ \mathsf{D}_i \}$ should be described by an appropriate moduli space of sheaves with some particular ``structure" (parabolic sheaves in the examples we have considered so far) which plays the role of the Hilbert space $\mathcal{H}^{BPS}_{ \{ \mathsf{D}_i \} }$. The enumerative problem then corresponds to the intersection theory of this moduli space, or in the most simple situation, to its Witten indices. In this paper we have refrained from considering this more general situation, and hope to return to it later on. For the time being is however instructive to mention a different point of view, according to which $k$-dimensional defects form a $k$-category in an Extended QFT \cite{Kapustin:2010ta}. In this language an $n$-dimensional field theory is,  roughly speaking, a certain functor $\mathsf{F}_n$ from the category of $n$-manifold with corners and some additional structure (in many applications in field theory one does not consider simply the category of topological manifolds, but is interested in manifolds with complex, k\"ahler, symplectic, etc... structures) to some $n$-category. For example when acting on an $n$-dimensional manifold $M_n$, the functor reproduces the partition function $\mathcal{Z} (M_n)$ of the field theory.  Similarly $\mathsf{F}_n (M_{n-1}) = \mathcal{H}$, is the Hilbert space of states. When considering defects of dimension $k$, one obtains a $k$-category $\mathsf{Def}^k$. If the defect is supported on a $k$ dimensional submanifold $D_k$, one considers a tubular neighbor of $D_k$, which locally looks like $\real^k \times S_r^{n-k-1} \times \real_+$. The $k$-category of defects $\mathsf{Def}^k$ is formally obtained in the limit where the radius $r$ of the sphere goes to zero as $\lim_{r \rightarrow 0} \ \mathsf{F}_n (S^{n-k-1}_r)$. From our perspective the partition function $\mathcal{Z} (M_n)$ corresponds to the generating function of ``volumes" of the moduli space of stable torsion free sheaves. In particular when $M_6$ is a toric Calabi-Yau, this partition function was studied in detail in the Coulomb branch in \cite{Cirafici:2008sn,Cirafici:2010bd}. The categorification of this partition function is a generating function of the motivic classes of the moduli spaces (studied in \cite{Cirafici:2011cd} in terms of quiver representation theory) corresponding to the space of states. In our example the $4$-category of divisor defects on a Calabi-Yau should correspond to the enumerative problem of torsion free sheaves with a parabolic structure along the divisor. The inclusion of further defects would imply the addition of extra conditions on this moduli space. On the other hand in the Extended QFT framework, $(k-m)$-dimensional defects supported on $D_k$ are labelled by $m$-morphisms of $\mathsf{Def}^k$. It would be fascinating to expand on this connection further, in particular to understand how the operation of adding extra structure to the moduli space of sheaves can be rephrased in the $k$-category framework.

\subsection{Generalizations}

The set of ideas discussed in this papers present several interesting further directions of investigation. The most immediate open problem is to study divisor defects in other topological field theories. We have briefly sketched how this could be done in an eight dimensional generalization of Donaldson-Thomas theory. Yet several other possibilities exists, by studying topological field theories defined on manifolds with certain structures. For example on could consider the eight dimensional theory defined in \cite{Baulieu:1997jx} on manifolds of $Spin(7)$ holonomy, or holomorphic Chern-Simons on a Calabi-Yau. Interestingly, several of these higher dimensional theories can be seen in certain limits as descriptions of topological M-theory \cite{Dijkgraaf:2004te}. This connection is rather intriguing and deserves further investigations.

On a different direction, we have considered explicitly only the case of the affine space. It is conceivable that our results generalize to arbitrary toric manifolds,  via gluing rules such as those developed in \cite{MNOP}. Note however that the same problem is still open in the case of four dimensional gauge theories with surface defects. The techniques discussed in this paper in the case of Donaldson-Thomas theory, also apply to four dimensional gauge theories. In particular one can construct partition functions of $\cN=2$ Yang-Mills theories on toric four manifolds with surface operators defined over compact divisors, by defining appropriate gluing rules. Work is in progress along this direction.

Finally, and perhaps more ambitiously, one would like to study the wall-crossing behavior of the Donaldson-Thomas invariants in the presence of a defect. On a generic toric threefold we already expect an intricate chamber structure, and possibly some of the techniques of \cite{Cirafici:2010bd,Cirafici:2011cd} can be applied. In particular it is conceivable that after a number of wall-crossings, our invariants could be related to the ones studied in \cite{toda}. We hope to report soon on these and other problems.

\section*{Acknowledgments}

The author was supported in part by the Funda\c{c}\~{a}o para a Ci\^{e}ncia e Tecnologia (FCT/Portugal) via the Ci\^{e}ncia2008 program and via the grant PTDC/MAT/119689/2010, and by the Center for Mathematical Analysis, Geometry and Dynamical Systems, a unit of the LARSyS laboratory.

\end{document}